\documentstyle[epsf]{mn}
\newif\ifAMStwofonts

\def\gtorder{\mathrel{\raise.3ex\hbox{$>$}\mkern-14mu
             \lower0.6ex\hbox{$\sim$}}}
\def\ltorder{\mathrel{\raise.3ex\hbox{$<$}\mkern-14mu
             \lower0.6ex\hbox{$\sim$}}}

\title{The redshift distribution of gravitational lenses revisited: Constraints on galaxy mass evolution}

\author[E.O. Ofek et al.]
       {Eran O. Ofek\thanks{e-mail: eran@wise.tau.ac.il}$^{1,2}$,  Hans-Walter Rix$^{2}$, Dan Maoz$^{1}$ \\
$^{1}$ School of Physics and Astronomy and Wise Observatory, Tel Aviv University, Tel Aviv 69978, Israel \\
$^{2}$ Max-Planck-Institut f\"{u}r Astronomie, K\"{o}nigstuhl 17, D-69117 Heidelberg, Germany}

\date{Accepted 2003 Apr 11th,
      Received 2002 Nov 14th}

\begin{document}

\maketitle

\begin{abstract}

The redshifts of lens galaxies in known gravitational lens systems
probe the volume distribution of lensing mass.
Following earlier work by Kochanek,
we re-derive the lens redshift probability distribution,
allowing for mass and number density evolution of the
lensing galaxies,
and apply this test to a much enlarged sample of lens systems.
From a literature survey of all known lenses,
we have selected an unbiased sample of 15 lenses
with complete redshift information.
For a flat Universe and no lens evolution,
we can only put an upper limit
on the cosmological constant of
$\Omega_{\Lambda}<0.89$ at the $95\%$ CL.
$\Omega_{\Lambda}\approx0.7$ and no evolution is
consistent with the data.
Allowing for evolution in
an $\Omega_{m}=0.3$, $\Omega_{\Lambda}=0.7$ cosmology, we find
that the best-fit evolution in $\sigma_{*}$
(i.e., the characteristic velocity dispersion in a Schechter-like function)
of early-type galaxies,
in the redshift range $z\sim0$ to $1$,
is $d\log{[\sigma_{*}(z)]}/dz=-0.10\pm0.06$.
This is consistent with no evolution and
implies that, at $95\%$ CL, $\sigma_{*}$ of
early-type galaxies at $z\sim1$ was at least $63\%$ of its current value.
Alternatively, if there is no mass evolution,
a present-day value of $\sigma_{*}>175$~km~s$^{-1}$
for elliptical galaxies is required ($95\%$ CL).

\end{abstract}


\section{Introduction}
\label{Introduction}

A number of recent measurements all point to a cosmology dominated
by dark energy.
The power spectrum of fluctuations in the
cosmic microwave background indicates a flat geometry
(e.g., Bennett et al. 2003),
while various measurements of the fraction of the closure density
in matter lead to consistent values of
$\Omega_{m}\approx0.3$ (e.g., Turner 2002).
Observations of high-redshift supernovae~Ia
(Riess et al. 1998; Perlmutter et al. 1999)
imply that the fraction of the closure density
in the form of a cosmological constant is $\Omega_{\Lambda}\approx0.7$.

However, some results obtained from gravitational lensing statistics
appear to be at odds with this picture.
The incidence of lensing among optically-selected quasars,
and their image-separation distribution, indicate
$\Omega_{\Lambda}<0.7$ at $95\%$ confidence level
(CL; Maoz \& Rix 1993; Kochanek 1996).
Similar studies of radio-selected sources 
(Falco, Kochanek, \& Munoz 1998) 
also find, $\Omega_{\Lambda}<0.73$
at the $95\%$ CL.
Chiba \& Yoshii (1999) have recalculated
the predicted lensing statistics,
using revised values for the galaxy luminosity function
parameters, and have argued that a Universe
with $\Omega_{\Lambda}=0.7$ is the most likely one.
Kochanek et al. (1998) defended their choice of
luminosity function parameters as the one that is most consistent
with the observed luminosities of lens galaxies.
Recently, Chae et al. (2002) derived constraints on cosmological parameters
from analysis of radio selected gravitational lenses
from the Cosmic Lens All Sky Survey (CLASS).
They find a best fit $\Omega_{\Lambda}=0.7\pm0.2$ and 
constrain $\Omega_{\Lambda}>0$ at $95\%$ CL,
assuming the most recent galaxy luminosity function parameters
from the 2dF survey (Madgwick et al. 2002) and the
Sloan Digital Sky Survey (SDSS; Blanton et al. 2001).
Note that, contrary to previous lensing statistics studies,
in Chae et al. (2002) the velocity dispersion
associated with a galaxy of given luminosity was left as a
free parameter, rather than being set by direct observations.
Their best fit velocity dispersions have relatively low values,
which result in low lensing cross sections. This may be the main
reason why their analysis favors
a relatively high value of $\Omega_{\Lambda}$.

Rix et al. (1994)
investigated the impact of hierarchical galaxy mergers on the statistics
of gravitational lensing of distant sources.
They showed that the {\it total}
number of multiple-image lenses
in a sample of potential sources
is quite insensitive to mergers, but
merging leads to a smaller mean separation of observed multiple images.
Since merging does not reduce drastically the expected lensing frequency,
it cannot make $\Lambda$-dominated cosmologies compatible with the
existing lensing observations (e.g., Maoz \& Rix 1993). 
Malhotra, Rhoads, \& Turner (1997) have suggested that
small-separation lensing statistics
can be reconciled with an $\Omega_{\Lambda}$-dominated Universe
by invoking dust in the lensing galaxies.
The excess number of lensed quasars would then be hidden by extinction.
Falco et al. (1999) found a mean extinction value of $\Delta{E_{B-V}}=0.04$~$(0.06)$
for lens galaxies in optically-selected (radio-selected) systems.
They tentatively concluded that the derived total
extinction in lenses accounts for the small differences in limits on
$\Omega_{\Lambda}$ found from lensing studies of radio selected
vs. optically-selected sources.

Another interesting statistic, that is the focus of this paper,
was suggested by Kochanek (1992),
and is based on the redshift of the lensing galaxy
in known lens systems.
Fukugita, Futamase, \& Kasai (1990) already showed that the
expected mean redshift of the deflector increases if $\Omega_{\Lambda}$
dominates over $\Omega_{m}$,
due to the large volume at high redshifts.
Kochanek (1992) obtained a formula for the probability
distribution of the lens-galaxy redshift given a source
(e.g., quasar) redshift and the critical radius
for lensing.
The critical radius for lensing was calculated assuming
galaxy lenses can be represented by singular isothermal spheres (SIS)
that are isolated (i.e., the lensing is not affected
by masses other than the lens galaxy).
The Kochanek method is applied by comparing
for each model (cosmology)
the redshift of the known lens, $z_{l}$,
with the redshift probability distribution of the lens galaxy
given the image separation
and the source redshift, $z_{s}$.
Using a sample of four lenses, Kochanek (1992)
found that the then-standard cosmologies
(without a cosmological constant) are 5-10 times more
probable than a flat cosmology with $\Omega_{\Lambda}\gtorder0.9$.
Since the image separation is taken into account as a prior,
this method is not affected by a bias resulting
from the fact that larger separation lenses
are more easily discovered.
This advantage allows, in principle, to use almost all the known lenses,
regardless of their method of discovery.
In \S\ref{Sample} we will discuss some exceptions.

Kochanek (1992) raised several questions and problems regarding
his method:
(i) How to deal with objects in which it is difficult
to detect the lens directly - should one trust a tentative lens redshift
based on absorption lines?
(ii) Is an unevolving galaxy mass distribution justified?
(iii) The lens system should not be selected
based on the lens-galaxy redshift.
For example, Q2237+0305 was discovered based on the quasar
emission lines present in the spectra of a nearby galaxy.
More examples of such objects, which may be found in the 2dF survey and
the SDSS (e.g., Mortlock \& Webster 2001),
are naturally biased toward lens galaxies at low redshift;
(iv) The method requires lenses
produced by isolated galaxies.
Otherwise, the image separations do not necessarily
represent the lens mass,
and indeed Kochanek
removed from his sample objects that are
influenced by clusters (e.g., Q0957+561).

The redshift distribution test was re-examined by
Helbig \& Kayser (1996), who compared the
redshifts of six lensing galaxies with
the probabilities predicted by
different cosmological models.
They assumed that there is no galaxy evolution,
and that lens galaxies could not be detected beyond
a certain magnitude $m_{lim}$.
For each tested cosmological model, they truncated the
redshift probability distribution beyond
the redshift at which the lens galaxy would
become too faint to have its redshift measured.
They concluded that
the method is not sensitive to cosmological parameters
and in the foreseeable future it would
probably not deliver interesting constraints.
Kochanek (1996) argued, however, that the insensitivity found by
Helbig \& Kayser (1996) was an artifact of their
statistics, due to the fact that they accounted
only for lenses with known redshifts,
while the fraction of the lenses with unmeasurable redshifts
was neglected.
The fraction of lens systems with measured redshifts
is smaller in a $\Lambda$-dominated Universe.
This is due to the higher lens redshifts for a given
source redshift, combined with a much larger luminosity distance to a
given redshift.
In order to account for the selection effect
introduced by the detectability of the lens galaxy,
Kochanek (1996) took into account also lens systems for which
the lens redshift was not measured.
Using a total of eight systems
(for four of which the lens redshift was available)
he obtained, for a flat Universe,
an upper limit of $\Omega_{\Lambda}<0.9$ at the $95\%$ CL.

Evolution of galaxy mass and number
can complicate or bias an analysis of the
lens redshift distribution.
For example, a scarcity of lenses at higher redshifts
may be due to evolution rather than to the smaller
volume of a low $\Omega_{\Lambda}$ cosmology.
In fact, there are observational indications for evolution
in the galaxy luminosity function (e.g., Cohen 2002),
and in the mass-to-light ratio (e.g., Keeton, Kochanek \& Falco 1998; van~Dokkum et al. 2001).
Evolution
of the lens luminosity function and $K$-corrections will also affect
the value of $m_{lim}$.
Cohen (2002) studied the galaxy evolution in
the HDF-North region and concluded that,
while the uncertainties are large, there is no sign of any
substantial increase in the total mass in stars between
$z=1.05$ and $z=0.25$.

Keeton (2002) has recently argued that 
lensing statistics predictions should be calibrated by counts of
distant galaxies at $z\sim0.5-1$
with properties similar to those of real lenses,
and not by the local density of galaxies.
Since distant-galaxy number counts depend themselves
on cosmological volume, when one calibrates the lensing
statistics by  the number counts of distant galaxies,
lensing statistics lose most
of their sensitivity to cosmology.
While in principle this is true, relating the observed
light from high redshift galaxies, and the light evolution,
to mass is not trivial.
The advantage of lensing is that it probes mass distribution
directly.
Therefore, it is still a valid approach to compare the observed
lensing statistics to the predictions of locally
calibrated models of lensing populations,
and to incorporate the mass and density evolution
into the models.
The mass evolution indicated by the lensing data can then
be compared to the luminosity evolution implied
by number counts, to gain insight both on cosmology
and on the relation between mass and light.

The purpose of this paper is to revisit the question of
the lens redshift distribution,
its dependence on cosmology, and on the evolution of the lens population,
for the larger sample of lensed quasars now available.
In \S\ref{For_LF} we rederive Kochanek's
formalism for the expected lens redshift distribution,
but allow for number and mass evolution of the lens population.
We also investigate the properties of the lens redshift distribution,
and its sensitivity to different parameters.
In order to define an unbiased sample,
in \S\ref{Sample} we conduct a literature survey
for known gravitational lens system and list them (in an Appendix)
with their basic parameters.
In \S\ref{Analysis} we constrain the cosmological
and mass-evolution parameter space.
We discuss the results
in \S\ref{Discussion},
and give a short summary in \S\ref{Summary}.

\section{Calculation of the Lens Redshift Distribution}
\label{For_LF}

\subsection{Formalism}
\label{Formalism}

The differential optical depth
to lensing per unit redshift is
\begin{equation}
\frac{d\tau}{dz}=n(\theta,z)(1+z)^{3} S \frac{cdt}{dz},
\label{opt_depth}
\end{equation}
where
$n(\theta,z)$ is the comoving number density of lenses
that have critical angular radius for lensing $\theta$ at redshift $z$,
$S$ is the cross section for lensing,
and $cdt/dz$ is the proper distance interval.

We assume that the lens galaxies can be represented by a SIS potential.
The SIS assumption is consistent with lens data,
galaxy dynamics, and the X-ray emission from
ellipticals (e.g., Fabbiano 1989; Rix et al. 1997; Treu \& Koopmans 2002).
Moreover, it is known that ellipticity in the lens galaxy mainly affects
the relative numbers of two- and four-image lenses 
but not their overall measures (e.g., Keeton, Kochanek, \& Seljak 1997).
The critical radius (Einstein radius),
and cross section, for lensing in a SIS potential are given by
\begin{equation}
\theta = 4\pi \left( \frac{\sigma}{c} \right)^{2} \frac{D_{ls}}{D_{s}} f_{E}^{2},
\label{Critical_radius}
\end{equation}
\begin{equation}
S = \pi \theta^{2} D_{l}^{2},
\label{Crros_section}
\end{equation}
where $D_{l}$, $D_{s}$, and $D_{ls}$ are the angular diameter distances
between  observer-lens, observer-source, and lens-source respectively
(e.g., Fukugita et al. 1992).
Note that in the SIS model, the separation between lensed images
is always $2\theta$.
Unless mentioned otherwise, we use
``filled-beam'' angular-diameter distances throughout this paper
(as opposed to ``empty-beam'', [Dyer \& Roeder 1972, 1973]).
$f_{E}$ is a free parameter that represents
the relation between dark halos and their stellar component,
as explained below.

The velocity dispersion $\sigma_{DM}$ of the mass distribution and
the observed stellar velocity dispersion $\sigma$ need not be the same.
Turner, Ostriker, \& Gott (1984) argued that if elliptical galaxies have
dark massive halos with more extended distributions than those of
the visible stars, this dark material must necessarily have
a greater velocity dispersion than the visible stars, and
a factor of $(3/2)^{1/2}$ is required to give an $r^{-2}$
density distribution (cf. Gott 1977).
Kochanek (1992) adopted a parameter
$f_{E}$ that relates the dark-matter velocity dispersion
$\sigma_{DM}$ and the stellar velocity dispersion $\sigma$:
\begin{equation}
\sigma_{DM}=f_{E}\sigma.
\label{f_E}
\end{equation}
Dynamical models
by Franx (1993), and Kochanek (1993, 1994)
demonstrated that the assumptions leading to the
$(3/2)^{1/2}$ factor were incorrect.
However, White \& Davis (1996) argued based on X-ray observations
that there is a strong indication that dark matter halos are dynamically hotter
than the luminous stars.
In our work, we have kept $f_{E}$ as a free parameter,
since it mimics the effects of:
(i) systematic errors in the velocity-dispersion measurements of galaxies;
(ii) large scale structures that change the typical separation
between images (e.g., Martel, Premadi, \& Matzner 2002; see \S\ref{Sample} below) - an increase of an arbitrary order $F$
in the velocity dispersion is equivalent to
an increase of $F^{2}$
in the typical separation $\theta$
(i.e., $\theta\propto\sigma^{2}$);
(iii) softened isothermal sphere potentials which tend to decrease the
typical image separations (e.g., Narayan \& Bartelmann 1996),
and could be represented by $f_{E}$ somewhat smaller than $1$.
Kochanek (1996) already noted that small core radii will lead to only slight
modifications of the tail of the lens redshift probability distribution.
Moreover, there is no evidence for cores in lensing galaxies
(e.g., Rusin \& Ma 2001).

Martel et al. (2002) compared the distribution of image separations
obtained from ray-tracing simulations in CDM models
(Premadi et al. 2001), with analytical calculations.
They found that the presence of the background matter tends to increase
the image separations produced by lensing galaxies.
However, they found this effect to be small,
of order of $20\%$ or less, and independent of
the cosmological parameters
(see also Bernstein \& Fischer 1999).
Bromley et al. (1998; see also Christlein 2000, Zabludoff \& Mulchaey 1998)
showed that the luminosity function of early type galaxies
depends on environment and that richer environments
tend to have a higher ratio of dwarf to giant galaxies than the field.
Keeton, Christlein, \& Zabludoff (2000) 
showed that this effect nearly (but not entirely)
cancels the effect of the background matter,
making the distribution of image separations essentially
independent of environment.
They predicted that lenses in groups have a mean image separation
that is $\sim0.2''$ smaller than that of lenses in the field.
It is possible that all these factors can affect the images separation by
up to $\pm20\%$.
This can be mimicked by introducing $(0.8)^{1/2}<f_{E}<(1.2)^{1/2}$.

In order to derive the lens redshift probability
we need the velocity-dispersion distribution function of galaxies.
Such a function is not yet available directly,
although it is expected that the SDSS and 2dF surveys,
combined with a good understanding of aperture
effects, will produce one soon.
Instead, we start with the Schechter (1976) function,
\begin{equation}
n(L,z) = n_{*}(z) \left(\frac{L}{L_{*}(z)} \right)^{\alpha} \exp{\left( -\frac{L}{L_{*}(z)} \right)} \frac{dL}{L_{*}(z)},
\label{SchechterFun}
\end{equation}
which provides a good description of the galaxy
luminosity distribution.
Using the SDSS data, Blanton et al. (2001) find that this function
fits the number density of SDSS galaxies
better than an alternative non-parametric model.
In order to relate the luminosity to velocity dispersion,
we use the Faber-Jackson relation
(Faber \& Jackson 1976; Kormendy \& Djorgovski 1989; Bernardi et al. 2001)
for early-type galaxies,
\begin{equation}
\frac{L}{L_{*}(z)} = \left( \frac{\sigma}{\sigma_{*}(z)} \right)^{\gamma}.
\label{FJ}
\end{equation}
A similar relation (i.e., a Tully-Fisher relation),
but with different $\sigma_{*}$ and $\gamma$,
is adopted for spiral galaxies.

From Equations~\ref{FJ} and~\ref{Critical_radius}
we obtain
\begin{equation}
\frac{L}{L_{*}(z)} = \left( \frac{\theta}{\theta_{*}(z)} \right)^{\gamma/2},
\frac{dL}{L_{*}(z)} = \frac{\gamma}{2} \left( \frac{\theta}{\theta_{*}(z)} \right)^{\gamma /2} \frac{d\theta}{\theta},
\label{FJ_theta}
\end{equation}
where
\begin{equation}
\theta_{*}(z) = 4\pi \left( \frac{\sigma_{*}(z)}{c} \right)^{2} \frac{D_{ls}}{D_{s}} f_{E}^{2}.
\end{equation}

In our rederivation of the lens redshift probability,
we allow for evolution in $n_{*}$, $L_{*}$,
and $\sigma_{*}$, with the following parametrization.
\begin{equation}
n_{*}(z)     = n_{*}  10^{Pz},
\label{n_star_z}
\end{equation}
\begin{equation}
L_{*}(z)     = L_{*}  10^{Qz},
\label{l_star_z}
\end{equation}
\begin{equation}
\sigma_{*}(z)     = \sigma_{*} 10^{Uz},
\label{s_star_z}
\end{equation}
where $P$, $Q$, and $U$ are constants.
$U$ represent an evolution in the characteristic $\sigma_{*}$, of
a Schechter-like ``mass'' function.
Note that the definitions of $P$ and $Q$ are somewhat different than the ones used by
Lin et~al (1999).
The definition of $Q$ is the same as the one used by Cohen (2002).
Throughout the paper we use $n_{*}$, $L_{*}$, $\sigma_{*}$, and $\theta_{*}$
to refer to the values at zero redshift, as opposed to $n_{*}(z)$, $L_{*}(z)$, $\sigma_{*}(z)$,
and $\theta_{*}(z)$.
Our parametrization is consistent with hierarchical galaxy merging.
Mergers change only the position of the galaxies
on the Faber-Jackson relation but not
the basic shape of the relation.
The scatter in the Faber-Jackson relation
(e.g., Kormendy \& Djorgovski 1989; Bernardi et al. 2001)
can be taken into account using a
Monte-Carlo simulation, as we discuss in \S\ref{Examples}.

Combining Equations~\ref{FJ_theta} and~\ref{SchechterFun},
to obtain $n(\theta,z)$, and inserting $n(\theta,z)$, along with the
evolution parametrization (Eq.~\ref{n_star_z}, \ref{l_star_z}, and \ref{s_star_z}) into Eq.~\ref{opt_depth}, we
obtain the optical depth per unit redshift
for a system with image separation $2\theta$ and source redshift $z_{s}$,
\begin{equation}
\begin{array}{ll}
\frac{d{\tau}}{dz} \left( \theta,z_{s} \right) =  & \tau_{N} 
             10^{z\left[-U\gamma(1+\alpha)+P \right]}  f_{E}^{2} \\
        & \times (1+z)^{3} \frac{D_{ls}}{D_{s}} D_{l}^{2} \frac{cdt}{dz} \left( \frac{\theta}{\theta_{*}} \right)^{\frac{1}{2}\gamma(1+\alpha)+1} \\
        & \times \exp{[- \left( \frac{\theta}{\theta_{*}} \right)^{\frac{1}{2}\gamma} 10^{-zU\gamma} ]},
\end{array}
\label{dtaudz_final}
\end{equation}
where the normalization,
\begin{equation}
\tau_{N} = 4\pi^{2}n_{*} \frac{\gamma}{2} \left(\frac{\sigma_{*}}{c} \right)^{2}.
\label{TauNorm}
\end{equation}
The lens light emission (as opposed to the lens mass)
does not affect lensing, and therefore
$Q$ does not appear in Equation~\ref{dtaudz_final}.
For a given lens system, the dependence of $d\tau /dz$ on $z$ gives the
relative probability of finding the lens at different $z$'s.

\subsection{Choices for the Lens Parameters}
\label{par_value}

Throughout this paper
we adopt the Schechter function parameters for the $b_{J}$-band, found by
Madgwick et al. (2002) from a sample of $75,000$ galaxies
in the 2dF survey with median redshift of $z\sim0.1$
(see also Blanton et al. 2001 for a similar result based on
a smaller sample using the SDSS).
We assume
$\alpha=-0.54\pm0.01$ for early-type galaxies (ellipticals and S0)
and $\alpha=-1.16\pm0.01$ as the average of late-type galaxies (weighted by their abundance).
The relative abundance of early-type to late-type galaxies is adopted from
the local density, as obtained by Madgwick et al. (2002),
while the relative abundance of ellipticals (E) to S0 is obtained
from Postman \& Geller (1984; E:S0=39:61).
We thus adopt
$n_{*}=1.46\times10^{-2}$; $0.61\times10^{-2}$; $0.39\times10^{-2}$~h$^{3}$~Mpc$^{-3}$
for spirals, S0s, and ellipticals, respectively.
The typical uncertainties in these number are of order $5\%$.
Note that these values are different from those used by
Kochanek (1992, 1996).
We do not incorporate evolution in $\alpha$, as
the redshift distribution test is not very sensitive to $\alpha$ (see \S\ref{Examples}).
A constant $\alpha$ is also justified by observations (Cohen 2002).
For the Faber-Jackson relations,
Bernardi et al. (2001) find $\gamma=4.00$, $3.91$, $3.95$, $3.92$
in the SDSS $g$, $r$, $i$, and $z$-bands, respectively,
with about $5\%$ error.
The spectral dependence of the slope $\gamma$,
is smaller than the typical scatter in the Faber-Jackson relation
($\sim40\%$ in a $\ln$-normal distribution; Bernardi et al. 2001).
We therefore adopt $\gamma=4.0$ for early-type galaxies.
For the Tully-Fisher relation of spirals, we assume
$\gamma=2.6$, following Kochanek (1992).
We take $\sigma_{*}$ from Fukugita \& Turner (1991),
based on de~Vaucouleurs \& Olson (1982), and
Efstathiou, Ellis, \& Peterson (1988):
$\sigma_{*}=225$~km~s$^{-1}$ for ellipticals;
$\sigma_{*}=206$~km~s$^{-1}$ for S0s;
and $\sigma_{*}=144$~km~s$^{-1}$ for spirals.
Note that
$M_{*}$ (the magnitude of a $\sigma_{*}$ galaxy)
given by Madgwick et al. (2001) and that given
by Efstathiou et al. (1988) agree well,
after applying the transformation
between the $b_{J}$ and $B_{T}$ magnitude systems (i.e., $b_{J}=B_{T}+0.29$).

Table~1 lists the default parameters we will assume in this paper. 
Unless mentioned otherwise,
$f_{E}=1$, $\Omega_{m}=0.3$, $\Omega_{\Lambda}=0.7$, $U=0$, $P=0$.
\begin{table}
\begin{tabular}{llll}
\hline
                                & Spiral                 & S0                   & Elliptical            \\
\hline
$\alpha$                        & $-1.16$                & $-0.54$              & $-0.54$               \\
$n_{*}$ [h$^{3}$~Mpc$^{-3}$]    & $1.46\times10^{-2}$    & $0.61\times10^{-2}$  & $0.39\times10^{-2}$   \\
$\gamma$                        & $2.6$                  & $4.0$                & $4.0$                 \\
$\sigma_{*}$  [km~s$^{-1}$]     & $144$                  & $206$                & $225$                 \\
\hline
\end{tabular}
\caption{The default parameters for the luminosity function and the Faber-Jackson relation.}
\end{table}

\subsection{What Can Be Tested by the Lens Redshift Distribution}
\label{Examples}

In this section we investigate the (generally complicated) behavior of
$d\tau/dz(\theta,z_{s})$
(Equation~\ref{dtaudz_final}) as a function
of the parameters involved in the problem.
We also study the effect of the scatter in the Faber-Jackson relation
on the lens redshift distribution.

\begin{figure*}
\centerline{\epsfxsize=170mm\epsfbox{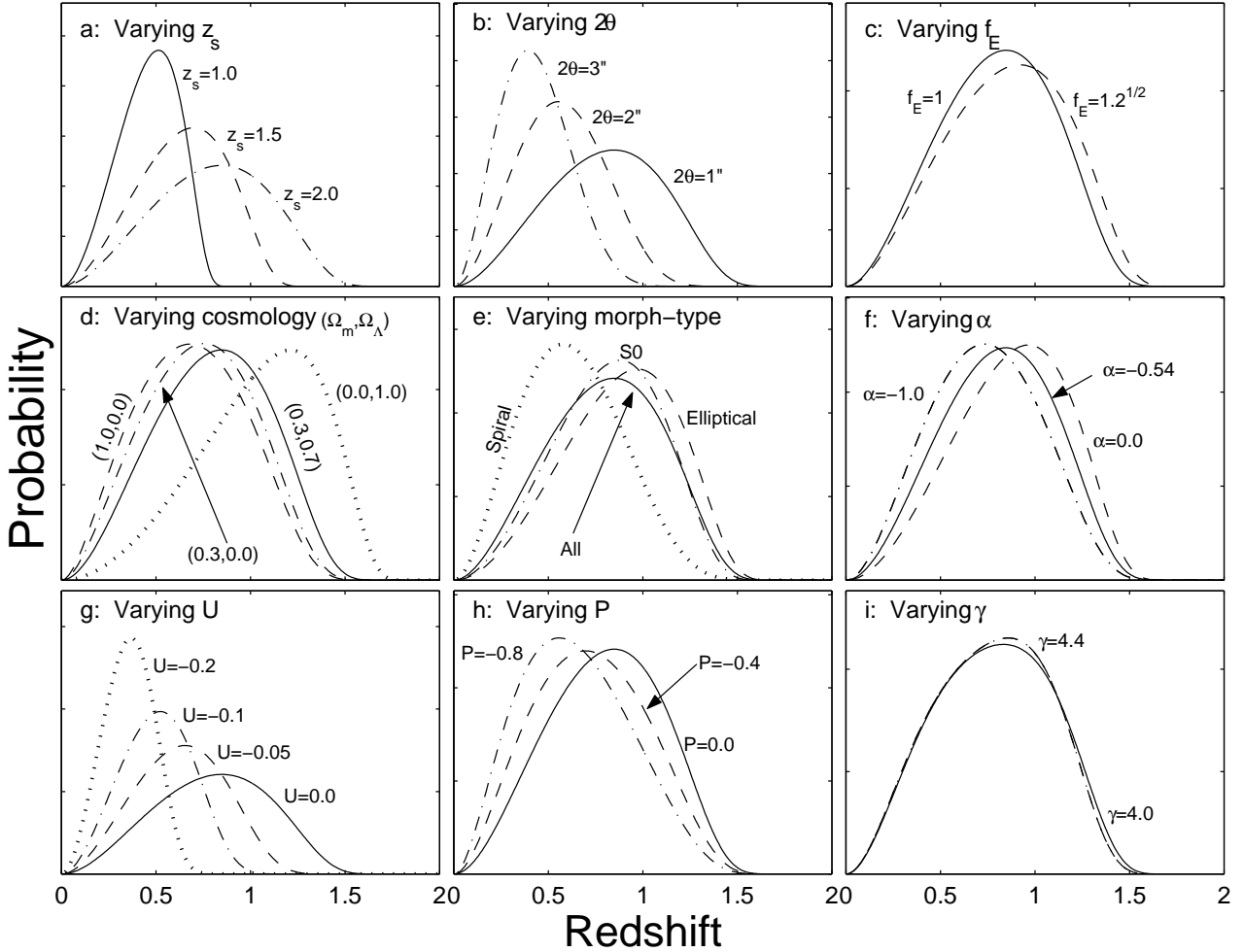}}
\caption{The lens redshift probability (normalized to one) density distribution
as a function of redshift for various choices of parameters.
The vertical axis is the probability in arbitrary units.
In all panels, unless otherwise noted, we use the default parameters:
$z_{s}=2$, $2\theta=1''$, $\Omega_{m}=0.3$, $\Omega_{\Lambda}=0.7$, $f_{E}=1$,
$U=0$, $P=0$, and the morphological mix with the parameters given in Table~1.}
\label{vari_parm}
\end{figure*}

Kochanek (1992) noted that the lens redshift distribution cuts off
more sharply at high redshift than the
total optical depth $d\tau/dz$
(i.e., Eq~\ref{dtaudz_final} integrated over all $\theta$'s),
because the critical radius constraint introduces an exponential term
from the Schechter function beyond the redshift at which an $L_{*}$
galaxy is required to produce the observed separation.
The sharp cutoff makes $d\tau / dz(\theta,z_{s})$ a powerful tool.
For example, the existence of relatively high
redshift lenses, even in a small sample,
can reject models in which such redshifts are beyond
the cutoff in the lens-redshift distribution.

Figure~\ref{vari_parm} shows the lens redshift probability
distribution for several different values of a given parameter,
while keeping the rest of the parameters fixed.
Unless otherwise noted,
we use the morphological mix of galaxies with the default parameters
described in \S\ref{par_value}, along with
$z_{s}=2$ and $2\theta=1''$.
Since it is not practical to scan all the parameter space
shown in Figure~\ref{vari_parm}, this figure can help us to
isolate the parameters that have the most profound effect
on the lens redshift distribution.

{\bf Evolution Parameters U and P:}
As seen in Fig~\ref{vari_parm}-g
the parameter $U$, which describes the evolution of the characteristic galaxy
velocity dispersion, has a considerable effect.
Assuming $P\sim0$, negative $U$ decreases the most probable redshift
of the lens $z$ distribution, and makes the distribution narrower.
This suggests that redshift evolution can impact the results
considerably,
and can therefore be constrained by the data.
The effect of the number density evolution parameter,
$P$ (see Fig~\ref{vari_parm}-h),
is somewhat smaller, and mainly changes the most probable
redshift, as opposed to $U$ that affects also
the cut-off redshift at which the
lens-redshift probability goes to zero.

{\bf Cosmology:}
From Figure~\ref{vari_parm}-d, it appears
that the lens redshift distribution test in a flat Universe is sensitive
to $\Omega_{\Lambda}$ mainly when $\Omega_{\Lambda}\gtorder0.8$.

{\bf Morphological type:}
It is well known (e.g., Maoz \& Rix 1993)
that almost all the contribution to
lensing optical depth comes from early-type galaxies.
Since the lensing galaxy type is not well constrained,
in the analysis we will take
it to be a mix of all the galaxy types
\footnote{Although several lens galaxies are
clearly disk systems (e.g., B1600+434),
even in those cases it is not clear whether they are S0's or later types.}.
As seen in Figure~\ref{vari_parm}-e, as long as early-type galaxies
dominate the lensing galaxy population,
there will be no dramatic effect on the final results.

{\bf ${\bf f_{E} }$:}
As discussed in \S\ref{For_LF},
the $f_{E}$ parameter (Figure~\ref{vari_parm}-c)
can mimic the uncertainties in the relation between image separation and mass,
or a systematic error in $\sigma_{*}$.
Although the effect of complex environments on the image separation
is not clear enough, it seems that in most cases it cannot change the image separation
by more than $20\%$ (see discussion in \S\ref{Introduction}).
Therefore, for reasonable values of $(0.8)^{1/2}<f_{E}<(1.2)^{1/2}$,
$f_{E}$ can somewhat affect the results of this work, but not as considerably
as the mass evolution or a large cosmological constant.
However, a systematic error of more than about $10\%$ in $\sigma_{*}$
can have some effect on the lens redshift distribution.

{\bf Angular diameter distance:}
We have tested the effect
of the angular diameter distance formulae
(filled-beam and empty-beam approximations)
on the lens redshift distribution (not shown in Fig~\ref{vari_parm}).
Using empty-beam approximation
somewhat decreases the typical lens redshift
relative to the filled-beam approximation,
but again, the effect is considerably smaller
than that of mass evolution
or of a large cosmological constant.

{\bf $ {\bf \gamma }$:}
Even a $10\%$ change in $\gamma$
does not cause significant variations in the lens redshift distribution.
Note that the effect of $\gamma$ in Figure~\ref{vari_parm}-i
is tested only for $U=0$, and its effect on the exponential
term of Eq.~\ref{dtaudz_final} is therefore limited.

{\bf ${\bf \alpha }$:}
The effect of changing $\alpha$ is shown in Figure~\ref{vari_parm}-f.
Madgwick et al. (2002) give a $2\%$ error on $\alpha$.
Changing $\alpha$ by $10\%$ (not shown) has a negligible effect,
even smaller then the effect of changing $\gamma$.
Setting $\alpha$ to a completely different value
(e.g., $\alpha=-1.1$; Efstathiou et al. 1988), somewhat shifts
the peak of the lens-redshift probability distribution.

\begin{figure}
\centerline{\epsfxsize=85mm\epsfbox{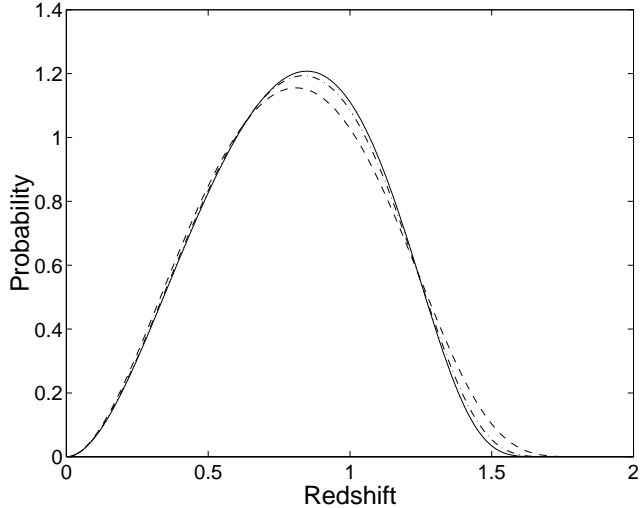}}
\caption {The lens-redshift probability distribution
for the default parameters, and $z_{s}=2$ and $2\theta=1''$ (solid line).
The dotted-dashed and dashed lines
were calculated using the same parameters,
but for the mean of $10,000$ Monte-Carlo realizations with
$\ln$-normal scatter introduced to $\sigma / \sigma_{*}$,
of $20\%$ and $40\%$ respectively.}
\label{Z_Dist_FJ_Scatter}
\end{figure}

An important issue that needs to be considered is the scatter in the
Faber-Jackson relation, which could alter the results.
The observed scatter in the Faber-Jackson relation is
about $40\%$ in a $\ln$-normal distribution (Bernardi et al. 2001).
To study the effect of such scatter,
we have performed Monte-Carlo simulations.
In each realization we perturbed
$\sigma / \sigma_{*}$ by multiplying it
by a factor randomly selected from a $\ln$-normal distribution
with a given standard deviation.
We performed two sets of Monte-Carlo simulations
with standard deviation of $20\%$, and $40\%$.
For each set we calculated the
lens-redshift probability, averaged over the realizations, as a function of redshift.
Figure~\ref{Z_Dist_FJ_Scatter} shows the mean effect of the
scatter in the Faber-Jackson relation on the
lens-redshift probability distribution.
The solid line is the unscattered lens-redshift probability distribution
as a function of redshift for the default parameters.
The dashed-dotted line is the same, but when $20\%$ scatter is introduced
to $\sigma / \sigma_{*}$, and the dashed line is for
a $40\%$ scatter.
It is clear that the Faber-Jackson scatter makes the distribution
somewhat broader.
However, it does not change the distribution significantly.
Note that using the lens redshift distribution function
based on a velocity dispersion distribution function would overcome
this problem.

Given the most probable values of the parameters,
the mass evolution $U$ has the most important effect on the
lens-redshift distribution.
This is not surprising since $U$ appears in
the exponential term of Equation~\ref{dtaudz_final}.
Next in importance are the cosmological constant,
the number density evolution, $P$, and $f_{E}$.
In what follows, we therefore investigate the parameters
found to be important to this problem, namely
$U$, $P$, $f_{E}$, and $\Omega_{\Lambda}$.

\section{Sample selection}
\label{Sample}

Using the lens redshift distribution
of known lenses to constrain cosmological parameters
and galaxy evolution requires
a sample free of selection effects and biases.
In Table~A1, we have compiled a list of $71$
galaxy lens systems
we are aware of (July 2002)
with their basic properties.
The main source for this list is the 
CASTLES database\footnote{http://cfa-www.harvard.edu/castles/index.html}
(Munoz et al. 1999)
that, at the time of writing, included $65$ systems.
We have added six additional systems discovered recently:
FIRST J100424.9+122922 (Lacy et al. 2002);
HS0810+2554 (Reimers et al. 2002);
PMNJ0134-0931 (Winn 2002b);
FIRST J0816+5003 (Leh{\' a}r et al. 2001);
HE0435-1223 (Wisotzki et al. 2002); PSS2322+1944 (Carilli et al. 2002).

There are several lens-system candidates that, although listed in the CASTLES database,
are of uncertain nature,
usually due to differences in the spectra of the supposed lensed
images or the non-detection of a lensing galaxy.
Rejection of these candidates based on the absence of a lensing galaxy
is problematic.
For example,
a lensing galaxy in a real lens system may be
undetected because it is distant and hence faint.
Systematically rejecting such systems would
bias the sample against detecting the effect of
a cosmological constant. 
We therefore reject lens candidates only
if their image spectra show pronounced differences.
Based on this criterion, Q0252-3249 (Morgan et al. 2000)
is probably not a real lens.
In any event, this object has $z_{s}=2.24$, and we will see below
that even if it were a real lens system it would not enter our sample.

As mentioned above, some lens systems have been discovered based on
the properties of the lens galaxy, thus
preferring low-redshift lenses.
On the other hand,
it is difficult to detect lensed quasars
for which the lens galaxy is nearby,
based on quasar selection.
After removing from our sample lens systems discovered
based on galaxy properties,
we therefore expect a bias against low-$z$ lenses.
In principle, we should correct for this bias
by modifying the lens-redshift distribution.
In practice, the probability for such lenses
at $z\ltorder0.1$
is very low anyway and we can safely ignore this correction.
Lenses rejected based on their discovery method
(i.e., lens-galaxy selected, as opposed to source selected)
are: Q2237+0305 (Huchra et al. 1985);
CFRS03.1077 (Crampton et al. 2002);
FIRST J0816+5003 (Leh{\' a}r et al. 2001).
We have also rejected all the lens candidates discovered
in the HST medium deep survey
(Ratnatunga, Griffiths, \& Ostrander 1999):
HST14176+5226;
HST12531-2914;
HST14164+5215;
HST15433+5352;
HST01247+0532;
HST01248+0531;
HST16302+8230;
HST16309+5215;
HST12369+6212;
HST18078+4600;
and in the Hubble Deep Fields:
\newline
HDFS2232509-603243 (Barkana, Blandford, \& Hogg 1999).
These lenses were selected based on their morphology
(e.g., the detection of a possible lensing galaxy)
and therefore, could introduce a bias.
Rejecting all these lens systems does not introduces any bias,
since we are effectively ignoring entire surveys.
Furthermore, only two of these systems have complete redshift
information, and as we will see below, their source redshift is too
high to be included in our sample.
In Column~10 of Table~A1 these objects
are marked ``L''.

There are several lens candidates
(listed as binary quasars in the CASTLES database)
whose nature is not clear
(e.g., Q1634+267, Peng et al. 1999; Q2345+007,
Small, Sargent, \& Steidel 1997, Green et al. 2002).
Even if these systems are real lenses, we reject them
from our sample, since their separation is too large
for galaxy lensing, and they could be affected
by cluster-mass objects.

Kochanek (1996) accounted for systems with missing redshifts
by considering
the probability that one cannot detect the
lens galaxy given the limiting magnitude to which
a system has been probed.
However, the lens galaxy
brightness depends on galaxy luminosity evolution
and this could complicate the analysis.
In order to avoid this problem we have defined a sample
that is source-redshift limited.
We have selected all systems with $z_{s}<2.1$ and image separations smaller than $4''$ (Sample~I).
This $z_{s}$ selection criterion has several advantages:
(i) Even for a de-Sitter Universe ($\Omega_{\Lambda}=1$), the most probable redshift is $<1.2$
and therefore we can probe the evolution factors within
a similar redshift range to that 
studied by optical surveys (e.g., Cohen 2002);
(ii) For $z_{s}<2.1$ we can expect that the redshifts for most
of the lenses have been measured, since they are not too high.
However, selecting a limit (as we have) that is based on the largest $z_{s}$
for which $z_{l}$ is measured for most systems, is somewhat biased.
For example, in a $\Lambda$-dominated Universe, the typical
$z_{l}$ will be higher and therefore the maximum $z_{s}$
(for which the sample is complete) will be lower.
To check for the robustness of the final results, we have
defined a second sample (Sample~II) for which the maximum $z_{s}$ is $1.7$.

The lens redshift information  in the samples is not complete.
There are three lens systems for which the lens redshift is not available yet.
These are J0158-4325 (Morgan et al. 1999), HE0435-1223 (Wisotzki et al. 2002),
and HS0810+2554 (Reimers et al. 2002).
All three systems were discovered recently, and spectral observations
{\it have not been attempted}.
However, the lens galaxies are detectable and are actually
quite bright ($R=19.4$, $i=18.1$, and $I\sim19$, respectively)
relative to other lens galaxies.
Therefore, dropping these systems from our sample will not introduce a bias.
For all but one of the lensing galaxies in samples~I and II,
the lens redshift has been measured spectroscopically.
The exception is FBQ0951+2635,
for which we have used the $BVRI$ photometry of 
Schechter et al. (1998), corrected for Galactic
extinction (Schlegel, Finkbeiner, \& Davis 1998),
to calculate a photometric redshift by
$\chi^{2}$ minimization relative to galaxy spectral templates
from Kinney et al. (1996).
We find a photometric redshift of $0.23$, $0.24$, and $0.28$,
for Sb, S0 and E0 galaxy types, respectively.
Kochanek et al. (2000)
showed that early-type field galaxies with
$0<z<1$, as represented by lens galaxies
in lensed quasar systems, lie on the same fundamental
plane as those in rich clusters at similar redshifts.
They used this to estimate the ``fundamental-plane redshift''
of lens galaxies by matching the lens galaxy
effective radius, magnitude and velocity dispersion
to the fundamental plane of rich clusters.
The Kochanek et al. (2000) fundamental-plane redshift
for FBQ0951+2635, although somewhat dependent on cosmology
through the luminosity distance,
is consistent ($z_{l}=0.21\pm0.03$)
with our photometric redshift.
Based on these estimates, we have adopted, $z_{l}=0.25$, for FBQ0951+2635.
In \S\ref{Analysis} we check for the robustness
of this assumption by changing the redshift of this object by $\pm0.05$.
The lens system B2045+265 was excluded from all the samples,
since its source redshift is most likely
in error (see Fassnacht et al. 1999).

The image separation criterion of $4''$ was introduced in order to
remove lenses that are influenced by complex environments (e.g., clusters).
The separation is used as prior information in calculating
the lens redshift probability for each system.
Therefore, even if we reject a lens that has a large separation,
only because its lens redshift is very low relative to the source redshift,
this selection will not bias the experiment.
On this basis we have removed Q0957+561, and RXJ0921+4529
from Samples~I and II.
To test for the robustness of this cut, we have
defined Sample~III, in which we have used sample~I
plus these two $\sim6''$ separation lenses.
Samples~I, II, and III contain 15, 11, and 17 sources, respectively.

For six lenses in sample~I for which an elliptical isothermal sphere 
plus shear model is available in the CASTLES database,
the critical radius ($\theta$) from the model is about
half the observed image separation
(or half the maximum separation in quadruple systems)
to $10\%$ accuracy or better.
Therefore, we adopt half the image separation
as representing the critical radius
in all systems.

\section{Analysis}
\label{Analysis}

\begin{figure*}
\centerline{\epsfxsize=170mm\epsfbox{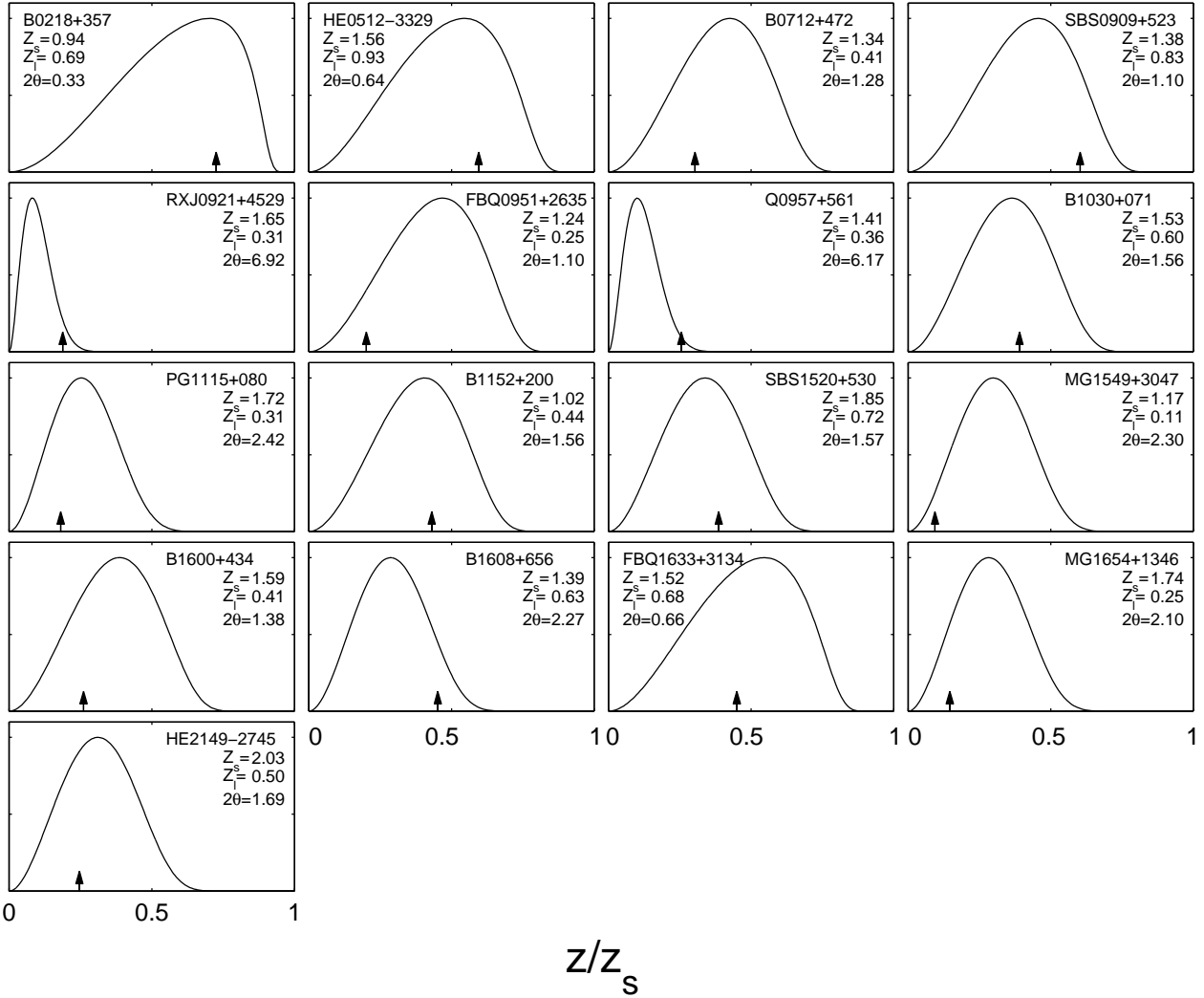}}
\caption {Lens redshift probability distributions calculated
for morphological mix of galaxies with
the default parameters (see \S\ref{par_value}),
for each of the $17$ lenses in samples~I, II, and III.
The vertical axis is the probability in arbitrary units.
In each panel the name of the lens along with its
source redshift ($z_{s}$), lens redshift ($z_{l}$) and
image separation in arcseconds ($2\theta$) are noted.
The observed lens redshift is marked by an arrow on the bottom axis.
}
\label{AllLenses}
\end{figure*}

Figure~\ref{AllLenses} shows the lens redshift probability distribution,
as calculated using
Equation~\ref{dtaudz_final} with the default parameters,
for each of the lenses in samples~I, II, and III.
In each panel the name of the lens along with its
source redshift ($z_{s}$), lens redshift ($z_{l}$), and image separation ($2\theta$) in arcseconds are listed.
The actual lens redshift is marked by an arrow on the bottom axis.

Our statistical analysis is based on the maximum likelihood technique.
For each lens system ($i$),
given the model (described by Equation~\ref{dtaudz_final}),
we calculate the probability density $P_{i}(z_{l} | z_{s}, \theta; \bf{X})$,
which is normalized to one,
where $\bf{X}$ are the model parameters (e.g., $\Omega_{\Lambda}$, $U$),
and $z_{s}$ and $\theta$ are priors.
The $\log$ likelihood ($\ln\cal{L}$)
as a function of $\bf{X}$ is given by
\begin{equation}
\ln{\cal L} = \sum_{i}{\ln{P_{i}}}.
\label{loglike}
\end{equation}
Testing one parameter at a time (i.e., $X_{j}$),
the confidence interval (CI)
within a given confidence limit (CL),
is the region found below the peak of the $\log$ likelihood
by $0.5$, $2.0$, and $3.3174$,
for $68\%$, $95\%$, and $99\%$ CLs, respectively (e.g., Press et al. 1992).

\begin{figure}
\centerline{\epsfxsize=85mm\epsfbox{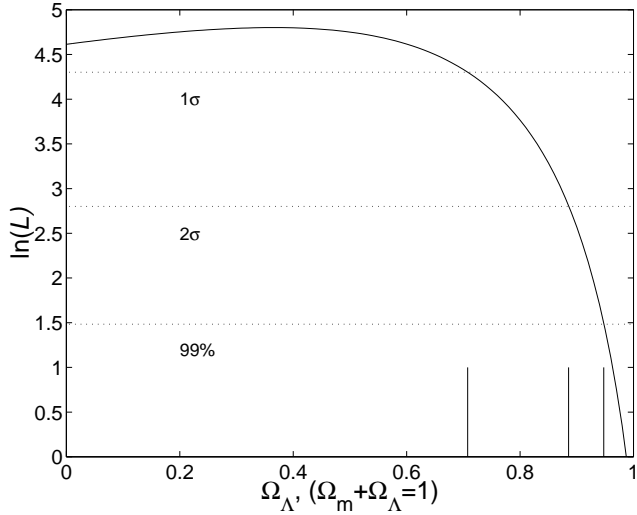}}
\caption {Likelihood plot for Sample~I.
The solid curve is the $\log$ of the likelihood as a function of $\Omega_{\Lambda}$ for
$\Omega_{m}+\Omega_{\Lambda}=1$. The rest of the parameters are the default ones.
The dotted lines are the $1\sigma$, $2\sigma$ and $99\%$ CLs. The vertical lines on the bottom axis mark the points where $\ln{\cal{L}}$ crosses the dotted lines.}
\label{ML_IV}
\end{figure}

Figure~\ref{ML_IV} shows $\ln\cal{L}$ for the full sample~I.
The dotted horizontal lines are the $68\%$, $95\%$, and $99\%$ CLs,
from top to bottom, respectively.
From this figure we see that the most likely cosmological model
(assuming no evolution) is $\Omega_{\Lambda}=0.36_{-0.70}^{+0.35}$,
with upper limits $\Omega_{\Lambda}<0.89$ at the $95\%$ CL,
and $\Omega_{\Lambda}<0.95$ at the $99\%$ CL.

The maximum-likelihood technique is not free of bias
(i.e., the most likely value could be biased; e.g., Efron 1982; Shao \& Tu 1995).
To test the results for bias and possible outliers,
we use the Quenouille-Tukey jackknife technique (e.g., Efron 1982; Efron \& Tibshirani 1993).
For a sample of $N$ ``observations'',
we remove one ``observation'' (i.e., one system) at a time,
and calculate $\ln\cal{L}$
as a function of $\Omega_{\Lambda}$,
for the $N-1$ remaining observations.
The jackknife estimate for the bias, $B$, in the parameter $\eta$
(i.e., $\Omega_{\Lambda}$) is given
by $B=(N-1)(<\eta_{i}> - \eta)$.
Where $\eta$ is estimated using $N$ observations,
and $\eta_{i}$ is the same,
but using $N-1$ observations at a time, where $i$ is the index
of observation dropped.
This bias should be subtracted from $\eta$ to
obtain the corrected $\eta$.
We find a bias of about
$\Delta\Omega_{\Lambda}\approx-0.2$.
Taking into account the bias,
the most probable $\Omega_{\Lambda}\approx0.6$ is well
within the $1\sigma$ error and therefore we will ignore the bias.

Sample~II favors a somewhat larger $\Omega_{\Lambda}=0.60_{-0.60}^{+0.27}$.
The higher $\Omega_{\Lambda}$ favored by sample~II
may be due to the $z_{s}$ cut we introduced (see \S\ref{Sample}).
However, the $\Omega_{\Lambda}$ values from the two samples
are consistent within
the uncertainty indicated by the likelihood ratios.
Sample~III favors an even higher cosmological constant,
$\Omega_{\Lambda}=0.77_{-0.22}^{+0.13}$
($1\sigma$ errors).
This result is not surprising, since cluster-affected lenses which have
large separation will tend to have 
lenses at redshifts higher than expected, assuming low $\Omega_{\Lambda}$
and ignoring the effect of the cluster.
This will increase the derived value of the cosmological constant.

As noted in \S\ref{Sample}, the redshift of the FBQ0951+2635
lens is based on a photometric redshift.
In order to test for the robustness
of this datum, we have recalculated $\ln\cal{L}$
when perturbing
the lens redshift
within the plausible range of redshifts
($z_{l}=0.25\pm0.05$; see \S\ref{Sample}), and have found that it has
little effect.

Given a model, the maximum-likelihood technique gives information about the
best fit parameters of the model.
However, the technique does not give information on the actual validity
(``goodness of fit'') of any model.
To this end, we have performed a set of Monte-Carlo simulations.
Given a set of model parameters,
in each simulation we used the source redshift ($z_{s}$) and image separation ($2\theta$)
of the objects in sample~I to generate a set of 1000 ``sample~I'' realizations.
Each realization contains 15 lens redshifts (like sample~I).
We then calculated the likelihood expectation value, and its distribution,
as a function of $\Omega_{\Lambda}$.
We find that the likelihood expectation value is
about $4.4\pm1.1$ ($1\sigma$ errors),
with some dependence on $\Omega_{\Lambda}$,
in good agreement with the maximum likelihood shown in Figure~\ref{ML_IV}.
We conclude that the lenses in sample~I
have redshifts consistent with the
probability function given by Equation~\ref{dtaudz_final},
and that our basic assumptions are probably valid.

\begin{figure}
\centerline{\epsfxsize=85mm\epsfbox{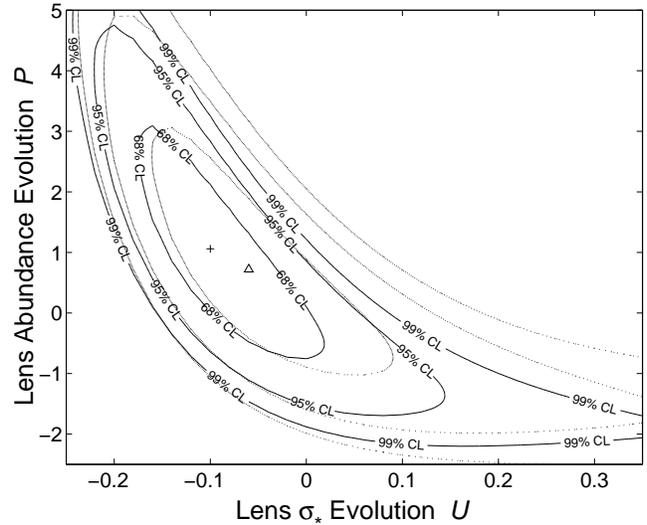}}
\caption {Likelihood contours
in the $U$-$P$ plane.
The solid curves are the, $1\sigma$, $2\sigma$ and $99\%$ CL lines
for sample~I, where the plus sign mark the position of
maximum likelihood.
The dotted curves are these contours for sample~II, for which
the triangle marks the position of the maximum.}
\label{ML_IV_II_UP}
\end{figure}

\begin{figure}
\centerline{\epsfxsize=85mm\epsfbox{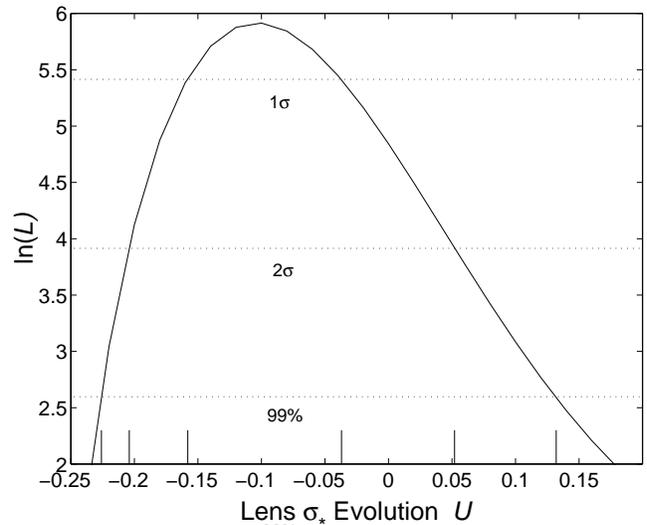}}
\caption {Marginalized likelihood as obtained
by integrating ${\cal L}(P,U)$ over $P$ (i.e., $\ln{\int{{\cal L}(P,U)}dP}$).
Symbols as in Fig~\ref{ML_IV}.}
\label{ML_IV_marU}
\end{figure}

\begin{figure}
\centerline{\epsfxsize=85mm\epsfbox{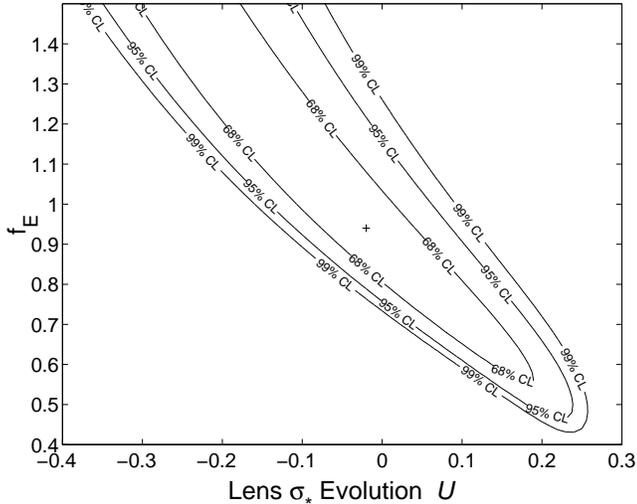}}
\caption {Likelihood contours
in the $U$-$f_{E}$ plane.
The solid curves are the, $1\sigma$, $2\sigma$ and $99\%$ CL lines
for sample~I, where the plus sign mark the position of
maximum likelihood.}
\label{ML_IV_Uf}
\end{figure}

Next, we set the cosmology to $\Omega_{m}=0.3$, $\Omega_{\Lambda}=0.7$,
as found by Bennett et al. (2003),
and maximize the likelihood for the evolution parameters.
Note that the evolution parameters $U$ and $P$
are applied to all morphological types.
As the cross section for spiral lensing
is considerably smaller than for ellipticals, $U$ and $P$
effectively trace the evolution in early-type galaxies.
Figure~\ref{ML_IV_II_UP} shows the likelihood contours
in the $U$-$P$ plane.
The solid curves are the $1\sigma$, $2\sigma$ and $99\%$ CL contours
(for two degree of freedom, see Press et al. 1992)
for sample~I, where the plus sign marks the position of
maximum likelihood.
The dotted curves mark the same contours for sample~II,
for which the triangle marks the position of the maximum likelihood.
We can see that both samples give consistent results.
As expected, the lens redshift distribution is a better
probe of $U$ than of $P$.
Marginalizing over $P$, Figure~\ref{ML_IV_marU} shows $\ln{\cal{L}}$
as obtained by integrating ${\cal L}(P,U)$
over $P$
(i.e., $\ln{\int{ {\cal L}(P,U)}dP}$).
In the absence of prior knowledge regarding $P$, we obtain
$U=-0.10_{-0.06,-0.10,-0.13}^{+0.06,+0.15,+0.23}$, where the errors
are for the $1\sigma$, $2\sigma$ and $99\%$-CL CIs, respectively.
Similarly, marginalizing over $U$,
we get $P=+0.7_{-1.2}^{+1.4}$ (not shown).
Finally, Figure~\ref{ML_IV_Uf} shows, for
$\Omega_{m}=0.3$, $\Omega_{\Lambda}=0.7$, and
no evolution in the number density of lenses
($P=0$), the likelihood contours in the $U$-$f_{E}$ plane.
Assuming that the dark-matter velocity dispersion 
and the stellar velocity dispersion are the same,
and that large scale structures do not change considerably
the image separations, then $f_{E}$ represent a
constant factor that multiplies the $\sigma_{*}$'s in
Table~1.
Thus, it mimics a systematic error in the ``canonical''
value of $\sigma_{*}$ used in this paper.
Assuming no evolution ($P=0$, $U=0$), we obtain
a best fit $f_{E}=0.90_{-0.07,-0.12,-0.15}^{+0.08,+0.19,+0.28}$
($1\sigma$, $2\sigma$ and $99\%$-CL errors).

\section{Discussion}
\label{Discussion}

The lens redshift distribution test, like other cosmological tests,
depends on assumptions, some of which may not be correct.
Based on the analysis we have conducted in \S\ref{Examples}, we would say
that our most troubling assumption is that image separation truly
represents the galaxy mass.
Large scale structure can influence the image separation, and therefore
we are probing some complex combination of galaxy mass and intervening
structures and not the mass of the galaxy on its own.
This is important as we use the ``mass function'' of galaxies to derive
the lens redshift distribution.
Our current knowledge suggests that in most cases,
the image separation is not affected by more than $20\%$ (see \S\ref{Formalism}) by large scale structure.

Extinction can influence somewhat the number statistics
of lenses, by lowering the observed number of optically selected lenses.
Falco et al. (1999) found a mean value of $\Delta{E_{B-V}}=0.05$ for lensed
galaxies.
The lens redshift distribution test is not influenced by dust,
due to the fact that known lens systems are used and their image separations
are taken as priors.
However, for calculating the lens-redshift probability,
we are using the observed mix of galaxy morphological types.
Since spirals are more dusty,
there could be a selection-induced lower fraction of spiral lenses relative to elliptical
lenses, and this effect would lower the
$\sigma_{*}$-evolution parameter, $U$, (or $\Omega_{\Lambda}$)
deduced by the lens redshift distribution method.
However, the lens population is
dominated by early-type galaxies by virtue of their larger mass.
As shown in Figure~\ref{vari_parm}, lowering the fraction
of spiral lenses will not affect the predictions dramatically.

We showed that the lens redshift distribution test is sensitive to high values
of the cosmological constant (e.g., $\Omega_{\Lambda}\gtorder0.8$).
Using the lens redshift distribution, and assuming
no evolution ($U=0$, $P=0$) and a flat Universe,
the current sample
puts an upper limit of $\Omega_{\Lambda}<0.89$ at
the $95\%$ CL, and $\Omega_{\Lambda}<0.95$ at the $99\%$ CL.
This is similar to the upper limit obtained by Kochanek (1996)
for a much smaller sample, using the same method.
Indeed, we find that the observations are consistent with the
``concordance'' value of $\Omega_{\Lambda}\approx0.7$.
Note, however, that this method
with the current sample is still not a sensitive probe of $\Omega_{\Lambda}$
for $\Omega_{\Lambda}\ltorder0.8$;
our results are equally consistent with $\Omega_{\Lambda}=0$.

We have shown that the lens redshift distribution is a
sensitive probe of the velocity-dispersion evolution of
early-type galaxies.
This arises from the exponential dependence on
the velocity-dispersion evolution parameter $U$ in Equation~\ref{dtaudz_final}.
Assuming $\Omega_{m}=0.3$, $\Omega_{\Lambda}=0.7$ 
and no prior knowledge of $U$ or $P$,
then the best fit values are:
$P=+0.7_{-1.2}^{+1.4}$ and
$U=-0.10\pm0.06$.
Our results apply to the redshift range $\sim0.1$ to $\sim0.9$.
The lower limit is set by the median redshift of galaxies used to obtain
the 2dF luminosity function (Madgwick et al. 2002)
and the Faber-Jackson relation (Bernardi et al. 2001).
The upper limit is set by the highest redshift lens in sample~I ($z=0.93$).
Our results are consistent with no evolution.
Positive $U$ (mass increases with redshift)
can probably be
excluded as physically improbable.
Our measurements are then useful for setting a lower limit
on this ``mass'' evolution parameter.
Negative $U$ would mean that the typical
characteristic velocity dispersion
(in a Schechter-like function) $\sigma_{*}$, was lower in the past.
Such evolution is expected in the hierarchical galaxy merging scenario,
as the typical galaxy was less massive in the past.
Our results imply that, at $95\%$ CL, $\sigma_{*}$ of
early-type galaxies at $z\sim1$ was at least $63\%$ of its current value.
As we have shown, the lens redshift distribution
does not strongly constrain the
number-density evolution $P$.
If we
take as a prior $P=-0.26$, as obtained by Im et al. (2002)
and supported by Schade et al. (1999)
from counts of early-type galaxies
to $z\approx1$, we see from Figure~\ref{ML_IV_II_UP} that
$U=-0.04_{-0.04,-0.07,-0.09}^{+0.05,+0.10,+0.13}$,
where the errors are the $1\sigma$, $2\sigma$ and $99\%$ CL errors,
respectively.
This places a more severe limit on the allowed evolution in $\sigma_{*}$;
at $95\%$ CL, $\sigma_{*}$ at $z\approx1$ was at least $77\%$
of its current value.

Most studies of galaxy evolution attempt to constrain the evolution
of the light, or of the mass-to-light ratio defined by
$(M/L)(z)=(M/L)_{0}10^{Tz}$,
rather than the mass evolution.
The luminosity evolution, number evolution,
and mass-to-light ratio evolution have been studied
in various works that produced
several estimates for the evolution parameters
$P$, $Q$, and $T$.
For example, Cohen (2002) studied the luminosity function
of galaxies in the HDF-North region,
in the redshift range $0$ to $1.5$.
She found that galaxies with strong absorption lines become
brighter with $z$ ($Q\sim0.6$ in all bands),
while galaxies with detectable emission lines show
a smaller change in $L_{*}$, with $Q\sim0.3$.
She also found that the comoving number density {\it and} stellar mass
in galaxies is approximately constant out to $z\sim1$,
and with more uncertainty, to $z\sim1.3$.
Keeton, Kochanek, \& Falco (1998) combined the photometry and lens modeling
of $17$ lens galaxies in the redshift range $0.1$ to $\sim1$.
They found that the lens galaxy $B$-band mass-to-light ratio decreases with redshift
with $T\approx-0.5\pm0.1$ (for $\Omega_{m}=0.1$).
In Table~2 we list these, and several other evolutionary studies of
early-type galaxies, and their best fit parameters.
The similar values of $-T$ and $Q$ found by most studies,
around $0.6$, suggest considerable luminosity
evolution with little mass evolution, consistent with our findings.
Note that the evolution parameters listed in Table~2
are determined for early-type galaxies in
different environments (field and clusters).
Interestingly, the evolution parameters are similar.

\begin{table}
\begin{tabular}{llll}
\hline
 Reference                        & $z_{max}$ & Parameters       \\
\hline
van~Dokkum et al. (1998)          & $0.83$    & $T\approx-0.45$  \\
Keeton, Kochanek, \& Falco (1998) & $\sim1$   & $T\approx-0.5$   \\
Lin et al. (1999)                 & $0.55$    & $P=-0.7\pm0.2$   \\
                                  &           & $Q=0.8\pm0.2$    \\
van~Dokkum et al. (2001)          & $0.55$    & $T=-0.59\pm0.15$ \\
Cohen (2002)                      & $\sim1$   & $Q\approx0.6$    \\
Rusin et al. (2002b)              & $\sim1$   & $T=-0.56\pm0.04$ \\
Im et al. (2002)                  & $\sim1$   & $Q=0.8\pm0.3$    \\
                                  &           & $P=-0.26\pm0.20$ \\
\hline
\end{tabular}
\caption{Evolutionary parameters
found using other methods.
$z_{max}$ is the maximum redshift for which the parameters were fitted.
Since the parameters are cosmology dependent,
we choose only works which assume a ``standard'' cosmology (i.e., $\Omega_{m}=0.3$, $\Omega_{\Lambda}=0.7$),
or cosmologies that have luminosity distances
that are not too different
(e.g., $q_{0}=0.1$ as used by Lin et al. 1999, or $\Omega_{0}=0.1$ used by Keeton et al. 1998).
Evolution in luminosity, number, and mass-to-light ratio
are parameterized by $Q$, $P$, and $T$, respectively.}
\end{table}

In a recent paper,  Davis, Huterer \& Krauss (2002)
use the CLASS and Jodrell-Very Large Array Astrometric Survey
lensing statistics to constrain the characteristic velocity dispersion
of elliptical galaxies.
This is done using the joint likelihood
for the number of lenses, the redshift distribution of lensing
galaxies, and the angular-separation distribution of lensed images.
Their analysis gives $138$~km~s$^{-1}\leq\sigma_{*}\leq 206$~km~s$^{-1}$
($95\%$ CL).
However, their lens-redshift distribution analysis
include only the six out of $12$ lenses in their sample
for which redshift information is available.
For a given image separation, lenses with higher redshift
have higher masses. Therefore, failing to take into account
the missing redshifts
will bias $\sigma_{*}$ toward lower values.
Furthermore, the ``canonical'' value of $\sigma_{*}$
(e.g., $225$~km~s$^{-1}$ for ellipticals; Fukugita \& Turner 1991)
is measured at redshift zero, while the lensing analysis constrains $\sigma_{*}$
at the mean redshift of the lenses.
Therefore, it is hard to tell
if the analysis by Chae et al. (2002) and Davis et al. (2002)
is detecting evolution in $\sigma_{*}$ or an error
in the old value of $\sigma_{*}$.
The mean lens-galaxy redshift of sample~I is $\overline{z_{l}}=0.54$.
Our best fit value of $U=-0.10$ can be mimicked
by a constant $\sigma_{*}$ that is about $88\%$ ($=10^{U\overline{z_{l}}}$)
of the local $\sigma_{*}$ we have used in this work.
However, we can see from Figure~\ref{ML_IV_Uf} that
a $\sigma_{*}$ that is $\ltorder80\%$ of the values
we have used in this paper (see Table~1),
can be rejected since it would require
positive evolution ($U$) at the $95\%$ CL.
Assuming no evolution ($U=0$, $P=0$),
we have found that the most probable $f_{E}=0.90_{-0.07,-0.12,-0.15}^{+0.08,+0.19,+0.28}$
($1\sigma$, $2\sigma$ and $99\%$-CL errors).
Therefore, a present-day value of $\sigma_{*}>175$~km~s$^{-1}$ for elliptical
galaxies is required ($95\%$ CL).
This value is in agreement with a constant velocity dispersion as found by
Chae et al. (2002), $\sigma_{*}=198^{+53}_{-37}$~km~s$^{-1}$ for ellipticals.
Note that Chae et al. (2002) use $\alpha=-1.0$,
as opposed to $\alpha=-0.54$.

Thus, a constant value of $\sigma_{*}\approx200$~km~s$^{-1}$
or a mildly evolving value that has approximately this average
would explain most of the current lensing statistics data
in an $\Omega_{\Lambda}=0.7$ cosmology.
Since lensing cross section is proportional to $\sigma^{4}$,
such a value of $\sigma_{*}$ may lower the lensing cross section
in lens surveys by a large enough amount to explain
the small number of observed lensed systems despite the
large volume associated with a cosmological constant.
At the same time our results show that such $\sigma_{*}$ is
still large enough to reproduce the observed redshift distribution
of known lenses.

\section{Summary}
\label{Summary}

We have compared the observed and predicted redshift distributions
of lensing galaxies.
Expanding upon the work of Kochanek (1992, 1996)
we have used a much enlarged observational sample
and incorporated possible mass and number density evolution of lens galaxies.
The lens-redshift distribution in a flat Universe is sensitive to the
large volume out to a given redshift in the presence of a large
cosmological constant (i.e., $\Omega_{\Lambda}\gtorder0.8$),
and especially to evolution of the characteristic velocity dispersion
of the lensing population.
The sensitivity of this test to velocity-dispersion evolution is unique,
as it probes the total mass of stars and dark matter within $\sim10$~kpc,
without any assumptions on a possible bias factor between
mass to light.
We have explored the robustness of our inferences
to various types of deviations from our basic assumptions,
such as errors in the luminosity-function parameters,
the lens model, and the scatter in the Faber-Jackson relation.
Unless there are gross errors in our knowledge of these parameters,
they have a small effect on our results.

There are now 71 lenses known in the literature
and we have compiled their basic properties (Appendix~\ref{LensTableApp}).
From these, we have selected a sample of 15 lenses
with separations smaller than $4''$ and source redshift $z_{s}<2.1$,
along with two other test samples.
Assuming no mass evolution ($0\ltorder z\ltorder 1$)
of early-type galaxies, we can put an upper limit of $\Omega_{\Lambda}<0.89$
at the $95\%$ CL.
On the other hand, assuming 
an $\Omega_{m}=0.3$, $\Omega_{\Lambda}=0.7$ cosmology
we find that the best fit
characteristic velocity dispersion ($\sigma_{*}$) and number evolution of early-type galaxies, in the range $z\sim0$ to $1$,
are
$d\log_{10}{\sigma_{*}(z)}/dz = -0.10\pm0.06$
and $d\log_{10}{n_{*}(z)}/dz = +0.7_{-1.2}^{+1.4}$.
Our findings are consistent with no evolution
in the early-type galaxy population and limit the allowed
increase of the characteristic $\sigma_{*}$
between $z\approx1$ and the present.
Finally, we have considered
the influence of changing the
estimate of the present-day value of $\sigma_{*}$.
We find that
a velocity dispersion that is $\ltorder80\%$ of the standard
value of $\sigma_{*}$ would require positive velocity-dispersion
evolution (i.e., galaxies were more massive in the past)
at the $95\%$ CL, and therefore can be rejected.
A value of $\sigma_{*}\approx200$~km~s$^{-1}$ for early-types,
with no evolution out to $z\approx1$, is consistent with our data.
Such a value of $\sigma_{*}$ could possibly also resolve the
discrepancy between the results of previous lensing statistics
studies and an $\Omega_{\Lambda}=0.7$ cosmology.

\section*{ACKNOWLEDGMENTS}
We thank Chris Kochanek,
Orly Gnat, Shay Zucker, Keren Sharon and Avishay Gal-Yam
for fruitful discussions.
We are grateful to the referee, D. Rusin, for his useful comments.
This work was supported by a grant from the
German Israeli Foundation for Scientific Research and Development.
EOO thanks
the Max-Planck-Institut f\"{u}r Astronomie
for its hospitality and
the Deutscher Akademischer Austauschdienst
for financial support.

\appendix

\section[]{List of known lenses}
\label{LensTableApp}

\begin{table*}
\centering
\begin{minipage}{185mm}
\begin{tabular}{lllllllllll}
\hline
Name              & R.A.         &  Dec.        & $z_{s}$ & $z_{l}$ & separation  &$mag_{lens}$ &$N_{im}$& Grade & Sample & Ref \\
(1)               & (2)          & (3)          & (4)     & (5)     & (6)            & (7)      & (8)   & (9)  & (10)   & (11)\\
\hline
Q0047-2808        & 00:49:41.89  & $-$27:52:25.7  & $3.595$ & $0.4845$& $2.7$   & $V=20.4$ & 4R     & A     &        & 2,32\\
HST01247+0532     & 01:24:44.4   & $+$03:52:00    &         &         & $2.17$  & $I=21.9$ & 2      & C     & L      & 3   \\
HST01248+0531     & 01:24:45.6   & $+$03:51:06    &         &         & $0.738$ & $I=22.6$ & 2      & C     & L      & 3   \\
Q0142-100         & 01:45:16.5   & $-$09:45:17    & $2.72$  & $0.49$  & $2.231$ & $R=19.4$ & 2      & A     &        & 32  \\
B0128+437         & 01:31:13.405 & $+$43:58:13.14 &         &         & $0.54$  &          & 4      & B     &        & 4   \\
PMNJ0134-0931     & 01:34:35.73  & $-$09:31:02.4  & $2.216$ & $0.76451$ &$0.7$  &          & 5      &       &        & 24,25,48\\
J0158-4325        & 01:58:41.44  & $-$43:25:04.20 & $1.29$  &         & $1.22$  & $R=19.4$ & 2      & A     &        & 5   \\
B0218+357         & 02:21:05.483 & $+$35:56:13.78 & $0.944$ & $0.685$ & $0.33$  & $I=20.1$ & 2R     & A     &I,II    & 51,52,62\\
HE0230-2130       & 02:32:33.1   & $-$21:17:26    & $2.162$ & $<0.6$  & $2.19$  & $V=19.9$ & 4      & A     &        & 6   \\
Q0252-3249        & 02:52:57.90  & $-$32:49:09.0  & $2.24$  &         & $2.1$   & $R>22.5$ & 2      & B     & N      & 8   \\
CFRS03.1077       & 03:02:32.65  & $+$00:06:00.2  & $2.941$ & $0.938$ & $2.1$   & $I=20.4$ & 2R     & B     & L      & 9   \\
MG0414+0534       & 04:14:37.73  & $+$05:34:44.3  & $2.64$  & $0.9584$& $2.12$  & $I=21.3$ & 4R     & A     &        & 10,14\\
HE0435-1223       & 04:38:14.90  & $-$12:17:14.4  & $1.689$ & $\sim0.4^{p}$&$2.6$&$i=18.1$ & 4      &       &        & 43  \\
HE0512-3329       & 05:14:10.78  & $-$33:26:22.50 & $1.565$ &$0.9313^{a}$&$0.644$&$I=17.6$ & 2      & A     &I,II    & 11  \\
B0712+472         & 07:16:03.58  & $+$47:08:50.0  & $1.339$ & $0.406^{b}$& $1.28$& $I=19.5$& 4      & A     &I,II    & 12  \\
B0739+367         & 07:42:51.2   & $+$36:34:43.7  &         &         & $0.54$  & $H=19.0$ & 2      & A     &        & 13  \\
MG0751+2716       & 07:51:41.46  & $+$27:16:31.35 & $3.200$ & $0.3502^{c}$& $0.7$& $I=21.2$& R      & A     &        & 14  \\
HS0810+2554       & 08:13:31.29  & $+$25:45:03.2  & $1.5$   &         & $0.25$  & $I\sim19$& 2      &       &        & 7   \\
FIRST J0816+5003  & 08:16:38.94  & $+$50:04:06.4  &         &         & $1.8$   & $R=19.2$ &        &       & L      & 28  \\
HS0818+1227       & 08:21:39.1   & $+$12:17:29    & $3.115$ & $0.39$  & $2.55$  & $I=18.8$ & 2      & A     &        & 63  \\
APM08279+5255     & 08:31:41.64  & $+$52:45:17.5  & $3.911$ &         & $0.378$ &          & 3      & A     &        & 15,16\\
SBS0909+523       & 09:13:01.05  & $+$52:59:28.83 & $1.377$ & $0.830$ & $1.10$  & $I=19.0$ & 2      & A     &I,II    & 27  \\
RXJ0911+0551      & 09:11:27.50  & $+$05:50:52.0  & $2.80$  & $0.77^{j}$& $3.25$ & $H=17.9$& 4      & A     &        & 30  \\
RXJ0921+4529      & 09:21:12.81  & $+$45:29:04.4  & $1.65$  & $0.31^{m}$& $6.92$& $I=20.2$ & 2      & B     &III     & 35  \\
FBQ0951+2635      & 09:51:22.57  & $+$26:35:14.1  & $1.24$  & $0.25^{n}$& $1.10$& $I=19.7$ & 2      & A     &I,II    & 53  \\
BRI0952-0115      & 09:55:00.01  & $-$01:30:05.0  & $4.50$  & $0.41^{o}$& $0.99$& $R=22.1$ & 2      & A     &        & 54  \\
Q0957+561         & 10:01:20.78  & $+$55:53:49.4  & $1.41$  & $0.36^{q}$& $6.17$& $I=17.1$ & 3R     & A     &III     & 55  \\
J100424.9+122922  & 10:04:24.872 & $+$12:29:22.39 & $2.65$  & $0.95^{r}$& $1.54$&          & 2      &       &        & 1   \\
LBQS1009-0252     & 10:12:15.71  & $-$03:07:02.0  & $2.74$  & $0.88^{s}$& $1.53$& $I=22.0$ & 2      & A     &        & 56  \\
Q1017-207         & 10:17:24.13  & $-$20:47:00.4  & $2.545$ & $1.085^{t}$& $0.849$& $I=21.8$& 2      & A     &        & 37  \\
FSC10214+4724     & 10:24:37.58  & $+$47:09:07.2  & $2.286$ & $0.914^{y}$& $1.59$& $I=20.4$& 2R     & A     &        & 57  \\
B1030+071         & 10:33:34.08  & $+$07:11:25.5  & $1.535$ & $0.599$ & $1.56$  & $I=20.2$ & 2      & A     &I,II    & 12  \\
HE1104-1805       & 11:06:33.45  & $-$18:21:24.2  & $2.32$  & $0.729$ & $3.19$  & $I=20.0$ & 2      & A     &        & 46  \\
PG1115+080        & 11:18:17.00  & $+$07:45:57.7  & $1.72$  & $0.311^{x}$&$2.42$& $I=18.9$ & 4      & A     &I       & 58  \\
B1127+385         & 11:30:00.13  & $+$38:12:03.1  &         & $>0.5^{d}$& $0.701$& $I=22.7$& 2      & A     &        & 17  \\
MG1131+0456       & 11:31:57.01  & $+$04:55:50.6  &         & $0.844^{f}$ & $2.1$& $I=21.2$& 2R     & A     &        & 19  \\
B1152+200         & 11:55:18.3   & $+$19:39:42.2  & $1.019$ & $0.439$  & $1.56$  & $I=19.6$& 2      & A     &I,II    & 36,45\\
Q1208+101         & 12:10:57.16  & $+$09:54:25.6  & $3.80$  & $1.1349^{h}$& $0.47$& $H>20$ & 2      & B     &        & 26  \\
HST12369+6212     & 12:36:49.0   & $+$62:12:22    &         &         & $1.166$ & $I=22.0$ & 2R     & C     & L      & 3   \\
HST12531-2914     & 12:53:06.70  & $-$29:14:30.0  &         & $0.63^{z}$& $1.09$  & $I=21.8$& 4     & A     & L      & 3   \\
B1359+154         & 14:01:35.55  & $+$15:13:25.6  & $3.235$ &$\approx1^{v}$& $1.7$&$I=22.7$& 6      & A     &        & 40  \\
HST14113+5211     & 14:11:19.60  & $+$52:11:29.0  & $2.811$ & $0.465$ & $2.26$  & $R=20.5$ & 4      & A     & L      & 27  \\
H1413+117         & 14:15:46.40  & $+$11:29:41.4  & $2.558$ &  $^{w}$ & $1.35$  & $H=18.6$ & 4      & A     &        & 59  \\
HST14164+5215     & 14:16:25.2   & $+$52:14:31    &         &         & $2.18$  & $I=20.8$ & 2      & C     & L      & 3   \\
HST14176+5226     & 14:17:36.51  & $+$52:26:40.0  & $3.40$  & $0.81$  & $3.25$  & $I=19.7$ & 4      & A     & L      & 3   \\
B1422+231         & 14:24:38.09  & $+$22:56:00.6  & $3.62$  & $0.339$ & $1.28$  & $I=19.6$ & 4R     & A     &        & 58  \\
SBS1520+530       & 15:21:44.83  & $+$52:54:48.6  & $1.855$ & $0.717^{ab}$&$1.568$& $I=20.2$& 2      & A     & I      & 31,32\\
HST15433+5352     & 15:43:20.9   & $+$53:51:52    &         &         & $1.176$ & $I=20.2$ & 2R     & C     & L      & 3   \\
MG1549+3047       & 15:49:12.37  & $+$30:47:16.6  & $1.17^{aa}$& $0.11$& $2.3$  & $I=16.7$ & R      & A     &I,II    & 47  \\
B1555+375         & 15:57:11.93  & $+$37:21:35.9  &         & $^{u}$  & $0.42$& $R=25$     & 4      & A     &I,II    & 38  \\
B1600+434         & 16:01:40.45  & $+$43:16:47.8  & $1.589$ & $0.4144$& $1.38$  & $I=20.6$ & 2      & A     &I,II    & 12  \\
B1608+656         & 16:09:13.96  & $+$65:32:29.0  & $1.394$ & $0.630$ & $2.27$  & $I=18.9$ & 4      & A     &I,II    & 39  \\
HST16302+8230     & 16:30:12.9   & $+$82:29:59    &         &         & $1.468$ & $I=23.7$ & 2R     & C     & L      & 3   \\
HST16309+5215     & 16:30:52.7   & $+$82:30:12    &         &         & $0.760$ & $I=20.2$ & 2R     & C     & L      & 3   \\
PMNJ1632-0033     & 16:32:57.68  & $-$00:33:02.05 & $3.424$ & $1.0^{e}$& $1.47$ & $I=23.1$ & 2R     & B     &        & 18  \\
FBQ1633+3134      & 16:33:48.99  & $+$31:34:11.90 & $1.52$  & $0.684^{k}$& $0.66$& $H=16.8$& 2      & B     &I,II    & 33  \\
MG1654+1346       & 16:54:41.83  & $+$13:46:22.0  & $1.74$  & $0.254$ & $2.1$  & $I=17.9$  & R      & A     &III     & 50  \\
HST18078+4600     & 18:07:46.7   & $+$45:59:56    &         &         & $0.912$ & $I=20.7$ & 2R     & C     & L      & 3   \\
PKS1830-211       & 18:33:39.94  & $-$21:03:39.7  & $2.51$  & $0.886$ & $0.99$  & $I=21.4$ & 2R     & A     &        & 60  \\
PMNJ1838-3427     & 18:38:28.8   & $-$34:27:33    & $2.78$  & $0.31^{g}$& $1.00$& $I=21.4$ & 2R     & A     &        & 20  \\
B1933+507         & 19:34:30.95  & $+$50:25:23.6  & $2.63$  & $0.755$ & $1.00$  & $I=20.2$ & 10     & A     &        & 41  \\
B1938+666         & 19:38:25.19  & $+$66:48:52.2  &         & $0.881$ & $1.0$   & $I=21.5$ & R      & A     &        & 19  \\
PMNJ2004-1349     & 20:04:07.09  & $-$13:49:31.1  &         &         & $1.13$  & $I=21.9$ & 2R     & A     &        & 21  \\
MG2016+112        & 20:19:18.15  & $+$11:27:08.3  & $3.27$  & $1.004$ & $3.26$  & $I=22.0$ & 3R     & A     &        & 42  \\
B2045+265         & 20:47:20.35  & $+$26:44:01.2  & $1.28$  & $0.8673$& $2.2$  & $I=21.2$  & 4      & A     &        & 44  \\
B2114+022         & 21:16:50.75  & $+$02:25:46.9  &         & $0.32^{i}$& $1.3$ & $I=18.6$ & 2+2    & B     &        & 29  \\
HE2149-2745       & 21:52:07.44  & $-$27:31:50.2  & $2.03$  & $0.50^{l}$&$1.69$ &$I=19.6$  & 2      & A     &I,II    & 34,32\\
\hline
\end{tabular}
\end{minipage}
\end{table*}
\begin{table*}
\centering
\begin{minipage}{185mm}
\begin{tabular}{lllllllllll}
\hline
Name              & R.A.         &  Dec.        & $z_{s}$ & $z_{l}$ & separation  &$mag_{lens}$ &$N_{im}$& Grade & Sample & Ref \\
(1)               & (2)          & (3)          & (4)     & (5)     & (6)            & (7)      & (8)    & (9)  & (10)   & (11)\\
\hline
HDFS2232509-603243& 22:32:50.9   & $-$60:32:43.0  &         &         & $0.9$   & $V=22$   & 4?R    & C     &  L     & 23  \\
Q2237+0305        & 22:40:30.34  & $+$03:21:28.8  & $1.695$ & $0.0394$& $1.82$  & $I=14.2$ & 4      & A     &  L     & 49  \\
B2319+052         & 23:21:40.8   & $+$05:27:36.4  &         & $0.624$ & $1.36$  & $H=18.2$ & 2      & A     &        & 22,27\\
PSS2322+1944      & 23:22:07.2   & $+$19:44:23    & $4.12$  &         & $1.5$   &          & 2      & B     &         & 61  \\
\hline
\end{tabular}
\caption{The columns are: (1) lens name; (2) R.A. [h:m:s], J2000.0; (3) Dec. [d:m:s], J2000.0;
(4) source redshift, $z_{s}$; (5) lens redshift, $z_{l}$; (6) image separation in arcseconds, $2\theta$; (7) lens magnitude;
(8) Number of images, R for ring;
(9) Grade for the likelihood that the object is a lens,
as adopted from the CASTLES database: A=I'd bet my life, B=I'd bet your life, 
and C=I'd bet your life and you should worry;
(10) sample: L-lens was discovered based on the lens-properties (rather than the source),
I- Sample~I in this paper, II- Sample~II in this paper, III-additional objects to those in Sample~I,
that are found in sample~III;
(11) References. \newline
List of references: \newline
$1$ - Lacy et al. (2002);
$2$ - Warren et al. (1996);
$3$ - Ratnatunga, Griffiths, \& Ostrander (1999);
$4$ - Phillips et al. (2000);
$5$ - Morgan et al. (1999);
$6$ - Wisotzki et al. (1999);
$7$ - Reimers et al. (2002);
$8$ - Morgan et al. (2000);
$9$ - Crampton et al. (2002);
$10$ - Falco, Leh{\' a}r, \& Shapiro (1997);
$11$ - Gregg et al. (2000);
$12$ - Fassnacht \& Cohen (1998);
$13$ - Marlow et al. (2001);
$14$ - Tonry \& Kochanek (1999);
$15$ - Ibata et al. (1999);
$16$ - Kobayashi et al. (2002);
$17$ - Koopmans et al. (1999);
$18$ - Winn et al. (2002a);
$19$ - Tonry \& Kochanek (2000);
$20$ - Winn et al. (2000);
$21$ - Winn et al. (2001);
$22$ - Rusin et al. (2001a);
$23$ - Barkana, Blandford, \& Hogg (1999);
$24$ - Winn et al. (2002b);
$25$ - Gregg et al. (2002);
$26$ - Siemiginowska et al. (1998);
$27$ - Lubin et al. (2000);
$28$ - Leh{\' a}r et al. (2001);
$29$ - Augusto et al. (2001);
$30$ - Kneib, Cohen, \& Hjorth (2000);
$31$ - Burud et al. (2002b);
$32$ - Kochanek et al. (2000);
$33$ - Morgan et al. (2001);
$34$ - Burud et al. (2002a);
$35$ - Mu\~{n}oz et al. (2001);
$36$ - Rusin et al. (2002a);
$37$ - Surdej et al. (1997);
$38$ - Marlow et al. (1999);
$39$ - Fassnacht et al. (1996);
$40$ - Rusin et al. (2001b);
$41$ - Sykes et al. (1998);
$42$ - Koopmans \& Treu (2002);
$43$ - Wisotzki et al. (2002) ;
$44$ - Fassnacht et al. (1999);
$45$ - Myers et al. (1999);
$46$ - Lidman et al. (2000);
$47$ - Leh{\' a}r et al. (1993);
$48$ - Hall et al. (2002);
$49$ - Huchra et al. (1985);
$50$ - Langston et al. (1989);
$51$ - O'Dea et al. (1992);
$52$ - Wiklind \& Combes (1995);
$53$ - Schechter et al. (1998);
$54$ - Leh{\' a}r et al. (2000);
$55$ - Walsh, Carswell, \& Weymann (1979);
$56$ - Hewett et al. (1994);
$57$ - Eisenhardt et al. (1996);
$58$ - Tonry (1998);
$59$ - Magain et al. (1988);
$60$ - Wiklind \& Combes (1996);
$61$ - Carilli et al. (2002);
$62$ - Cohen, Lawrence, \& Blandford (2002);
$63$ - Hagen \& Reimers (2000)}
Notes: \newline
$a$ - Based on absorption line spectrum (Gregg et al. 2000). \newline
$b$ - Associated with a group at $z=0.29$ and virial mass
of $\sim300$~km~s$^{-1}$ (Fassnacht \& Lubin 2002). \newline
$c$ - Associated with a small group and possibly a second group at $z=0.52$ (Tonry \& Kochanek 1999). \newline
$d$ - Two late-type lensing galaxies (Koopmans et al. 1999). \newline
$e$ - Redshift is based on the Kochanek et al. (2000) method in which one requires that the
galaxy properties be consistent with the passively evolving fundamental plane of early-type galaxies (Winn et al. 2002a). \newline
$f$ - Evidence for a foreground group of galaxies at $z=0.34$ (Tonry \& Kochanek 2000). \newline
$g$ - Redshift is based on the Kochanek et al. (2000) fundamental-plane method,
plus the possible identification of [OIII] in emission (Winn et al. 2002b). \newline
$h$ - The most probable redshift is of an Mg II absorber at $z=1.1349$ (Siemiginowska et al. 1998). \newline
$i$ - Complex lens, probably by two galaxies at $z=0.3157$ and $z=0.5883$ (Augusto et al. 2001). \newline
$j$ - Massive cluster at $z=0.769$ with $\sigma\sim840$~km~s$^{-1}$ (Kneib, Cohen, \& Hjorth 2000). \newline
$k$ - Spectral observations reveal a rich metal-line absorption system consisting of a strong Mg~II doublet and associated Fe~I and Fe~II absorption features, all at an intervening redshift of $z=0.684$, suggestive of a lensing galaxy (Morgan et al. 2001). \newline
$l$ - By cross correlating the spectrum with galaxy templates, Burud et al. (2002a)
obtain a tentative redshift estimate of $z=0.50$, in agreement with the fundamental-plane method estimate (Kochanek et al. 2000). \newline
$m$ - Possibly a member of an X-ray cluster ($z=0.32$) centered on the field (Mu\~{n}oz et al. 2001). \newline
$n$ - Based on photometric redshift (this paper). The fundamental-plane method (Kochanek et al. 2000) gives $z_{l}=0.21\pm0.03$. \newline
$o$ - $z_{l}=0.41\pm0.05$ is based on Kochanek et al. (2000) fundamental-plane method. \newline
$p$ - Based on lensing galaxy colors $z=0.4\pm0.1$ (Wisotzki et al. 2002). \newline
$q$ - Cluster at $z=0.36$ (e.g., Chae 1999). \newline
$r$ - Based on the probable identification of the lens galaxy $4000$\AA break in the quasar spectra
(Lacy et al. 2002). \newline
$s$ - Based on the fundamental-plane method (Kochanek et al. 2000) $z=0.88\pm0.07$. \newline
$t$ - Absorption lines at $z=2.344$ and MgII absorption at $z=1.085$ (Surdej et al. 1997). Kochanek et al. (2000) based on the fundamental-plane method give $z=0.78\pm0.07$. Therefore, we have adopted $z=1.085$. \newline
$u$ - Marlow et al. (1999) argue that the lens galaxy is sub-$L_{*}$, so it has $z>0.5$, or is highly reddened. \newline
$v$ - Lensing by a compact group of three $z\approx1$ galaxies (Rusin et al. 2001b). \newline
$w$ - Possible cluster at $z\sim1.7$ (Kneib et al. 1998). \newline
$x$ - Keeton \& Kochanek (1997) find that the lens time-delay cannot be fit by an isolated lens galaxy, but that it can be well fitted by including a perturbation from the nearby group of galaxies. Tonry (1998) measures the group velocity dispersion of $330$~km~s$^{-1}$. \newline
$y$ - Eisenhardt et al. (1996) estimate  the lens-galaxy redshift as $z=0.9\pm0.3$. Goodrich et al. (1996) find strong evidence for MgII absorption at $z=1.316$ in the spectrum of the QSO and weaker evidence for a possible continuum and absorption system at $z=0.893$. Lacy, Rawlings, \& Serjeant (1998) identify an H$\alpha$ emission line with $z=0.914$. The Kochanek et al. (2000) fundamental plane method gives $z=0.78\pm0.05$. \newline
$z$ - Based on the Kochanek et al. (2000) fundamental plane method. \newline
$aa$ - L. Koopmans and T. Treu, private communication (radio ring detected in radio). \newline
$ab$ - Excess of galaxies at $z\sim0.9$ (Faure et al. 2002).
\end{minipage}
\end{table*}


\begin{thebibliography}{}

\bibitem[Augusto et al.(2001)]{2001MNRAS.326.1007A} Augusto, P.~et al.\ 
2001, MNRAS, 326, 1007

\bibitem{Bernardi2001} Bernardi, M., et al., 2001, submitted to AJ, astro-ph/0110344

\bibitem[Barkana, Blandford, \& Hogg(1999)]{1999ApJ...513L..91B} Barkana, 
R., Blandford, R., \& Hogg, D.~W.\ 1999, ApJL, 513, L91

\bibitem[Bernstein \& Fischer(1999)]{1999AJ....118...14B} Bernstein, G.~\& 
Fischer, P.\ 1999, AJ, 118, 14

\bibitem[Blanton et al.(2001)]{2001AJ....121.2358B} Blanton, M.~R.~et al.\ 
2001, AJ, 121, 2358

\bibitem[Bromley, Press, Lin, \& Kirshner(1998)]{1998ApJ...505...25B} 
Bromley, B.~C., Press, W.~H., Lin, H., \& Kirshner, R.~P.\ 1998, ApJ, 505, 
25

\bibitem[Burud et al.(2002a)]{2002A&A...383...71B} Burud, I.~et al.\ 2002a, 
A\&A, 383, 71

\bibitem[Burud et al.(2002b)]{2002A&A...391..481B} Burud, I.~et al.\ 2002b, A\&A, 391, 481

\bibitem[Carilli et al.(2002)]{2002ApJ...575..145C} Carilli, C.~L.~et al.\ 
2002, ApJ, 575, 145

\bibitem[Chae(1999)]{1999ApJ...524..582C} Chae, K.\ 1999, ApJ, 524, 582

\bibitem[Chae et al.(2002)]{2002PhRvL..89o1301C} Chae, K.-H.~et al.\ 2002, 
Physical Review Letters, 89, 151301

\bibitem{CY99} Chiba, M., Yoshii, Y. 1999, ApJ, 510, 42

\bibitem[Christlein(2000)]{2000ApJ...545..145C} Christlein, D.\ 2000, ApJ, 
545, 145

\bibitem[Cohen(2002)]{2002ApJ...567..672C} Cohen, J.~G.\ 2002, ApJ, 567, 
672

\bibitem{CLB2002} Cohen, J.G., Lawrence, C.R., Blandford, R.D., 2002, Accepted for ApJ, astro-ph/0209457

\bibitem[Crampton et al.(2002)]{2002ApJ...570...86C} Crampton, D., Schade, 
D., Hammer, F., Matzkin, A., Lilly, S.~J., \& Le F{\` e}vre, O.\ 2002, 
ApJ, 570, 86

\bibitem[Faure et al.(2002)]{2002A&A...386...69F} Faure, C., Courbin, F., 
Kneib, J.~P., Alloin, D., Bolzonella, M., \& Burud, I.\ 2002, A\&A, 386, 69

\bibitem{DHK2002} Davis, A.N., Huterer, D., Krauss, L.M., 2002, submitted to MNRAS, astro-ph/0210494


\bibitem[Dyer \& Roeder(1972)]{1972ApJ...174L.115D} Dyer, C.~C.~\& Roeder, 
R.~C.\ 1972, ApJL, 174, L115

\bibitem[Dyer \& Roeder(1973)]{1973ApJ...180L..31D} Dyer, C.~C.~\& Roeder, 
R.~C.\ 1973, ApJL, 180, L31 

\bibitem[Eisenhardt et al.(1996)]{1996ApJ...461...72E} Eisenhardt, P.~R., 
Armus, L., Hogg, D.~W., Soifer, B.~T., Neugebauer, G., \& Werner, M.~W.\ 
1996, ApJ, 461, 72

\bibitem{Efron1982} Efron, B., 1982, The Jackknife, the Bootstrap and Other Resampling Plans, The Society for Industrial and Applied Mathematics

\bibitem{ET1993} Efron, B., Tibshirani, R.J., 1993, An introduction to the bootstrap, Monographs on statistics and applied probability 57, Chapman \& Hall

\bibitem[Efstathiou, Ellis, \& Peterson(1988)]{1988MNRAS.232..431E} 
Efstathiou, G., Ellis, R.~S., \& Peterson, B.~A.\ 1988, MNRAS, 232, 431

\bibitem[Fabbiano(1989)]{1989ARA&A..27...87F} Fabbiano, G.\ 1989, ARA\&A, 
27, 87

\bibitem[Faber \& Jackson(1976)]{1976ApJ...204..668F} Faber, S.~M.~\& 
Jackson, R.~E.\ 1976, ApJ, 204, 668

\bibitem[Falco, Lehar, \& Shapiro(1997)]{1997AJ....113..540F} Falco, E.~E., Leh{\' a}r, J., \& Shapiro, I.~I.\ 1997, AJ, 113, 540

\bibitem[Falco, Kochanek, \& Munoz(1998)]{1998ApJ...494...47F} Falco, 
E.~E., Kochanek, C.~S., \& Munoz, J.~A.\ 1998, ApJ, 494, 47

\bibitem[Falco et al.(1999)]{1999ApJ...523..617F} Falco, E.~E.~et al.\ 
1999, ApJ, 523, 617

\bibitem[Fassnacht et al.(1996)]{1996ApJ...460L.103F} Fassnacht, C.~D., 
Womble, D.~S., Neugebauer, G., Browne, I.~W.~A., Readhead, A.~C.~S., 
Matthews, K., \& Pearson, T.~J.\ 1996, ApJL, 460, L103

\bibitem[Fassnacht \& Cohen(1998)]{1998AJ....115..377F} Fassnacht, C.~D.~\& 
Cohen, J.~G.\ 1998, AJ, 115, 377

\bibitem[Fassnacht et al.(1999)]{1999AJ....117..658F} Fassnacht, C.~D.~et 
al.\ 1999, AJ, 117, 658

\bibitem[Fassnacht \& Lubin(2002)]{2002AJ....123..627F} Fassnacht, C.~D.~\& 
Lubin, L.~M.\ 2002, AJ, 123, 627

\bibitem[Faure et al.(2002)]{2002A&A...386...69F} Faure, C., Courbin, F., 
Kneib, J.~P., Alloin, D., Bolzonella, M., \& Burud, I.\ 2002, A\&A, 386, 69

\bibitem[Franx(1993)]{1993ApJ...407L...5F} Franx, M.\ 1993, ApJL, 407, L5

\bibitem[Fukugita, Futamase, \& Kasai(1990)]{1990MNRAS.246P..24F} Fukugita, 
M., Futamase, T., \& Kasai, M.\ 1990, MNRAS, 246, 24P

\bibitem[Fukugita \& Turner(1991)]{1991MNRAS.253...99F} Fukugita, M.~\& 
Turner, E.~L.\ 1991, MNRAS, 253, 99

\bibitem[Fukugita, Futamase, Kasai, \& Turner(1992)]{1992ApJ...393....3F} 
Fukugita, M., Futamase, T., Kasai, M., \& Turner, E.~L.\ 1992, ApJ, 393, 3

\bibitem[Goodrich et al.(1996)]{1996ApJ...456L...9G} Goodrich, R.~W., 
Miller, J.~S., Martel, A., Cohen, M.~H., Tran, H.~D., Ogle, P.~M., \& 
Vermeulen, R.~C.\ 1996, ApJL, 456, L9

\bibitem[Gott(1977)]{1977ARA&A..15..235G} Gott, J.~R.\ 1977, ARA\&A, 15, 235

\bibitem[Green et al.(2002)]{2002ApJ...571..721G} Green, P.~J.~et al.\ 
2002, ApJ, 571, 721

\bibitem[Gregg et al.(2000)]{2000AJ....119.2535G} Gregg, M.~D., Wisotzki, 
L., Becker, R.~H., Maza, J.~;, Schechter, P.~L., White, R.~L., Brotherton, 
M.~S., \& Winn, J.~N.\ 2000, AJ, 119, 2535

\bibitem[Gregg et al.(2002)]{2002ApJ...564..133G} Gregg, M.~D., Lacy, M., 
White, R.~L., Glikman, E., Helfand, D., Becker, R.~H., \& Brotherton, 
M.~S.\ 2002, ApJ, 564, 133

\bibitem[Hagen \& Reimers(2000)]{2000A&A...357L..29H} Hagen, H.-J.~\& 
Reimers, D.\ 2000, A\&A, 357, L29

\bibitem{Hall2002} Hall, P.B., et al. 2002, accepted to ApJL, astro-ph/0207317


\bibitem{HK1996} Helbig, P., Kayser, R., 1996, A\&A, 308, 359

\bibitem[Hewett et al.(1994)]{1994AJ....108.1534H} Hewett, P.~C., Irwin, 
M.~J., Foltz, C.~B., Harding, M.~E., Corrigan, R.~T., Webster, R.~L., \& 
Dinshaw, N.\ 1994, AJ, 108, 1534

\bibitem[Huchra et al.(1985)]{1985AJ.....90..691H} Huchra, J., Gorenstein, 
M., Kent, S., Shapiro, I., Smith, G., Horine, E., \& Perley, R.\ 1985, AJ, 
90, 691

\bibitem[Ibata et al.(1999)]{1999AJ....118.1922I} Ibata, R.~A., Lewis, 
G.~F., Irwin, M.~J., Leh{\' a}r, J., \& Totten, E.~J.\ 1999, AJ, 118, 1922

\bibitem[Im et al.(2002)]{2002ApJ...571..136I} Im, M.~et al.\ 2002, ApJ, 571, 136

\bibitem[Keeton \& Kochanek(1997)]{1997ApJ...487...42K} Keeton, C.~R.~\& 
Kochanek, C.~S.\ 1997, ApJ, 487, 42

\bibitem[Keeton, Kochanek, \& Seljak(1997)]{1997ApJ...482..604K} Keeton, 
C.~R., Kochanek, C.~S., \& Seljak, U.\ 1997, ApJ, 482, 604

\bibitem[Keeton, Kochanek, \& Falco(1998)]{1998ApJ...509..561K} Keeton, 
C.~R., Kochanek, C.~S., \& Falco, E.~E.\ 1998, ApJ, 509, 561

\bibitem[Keeton, Christlein, \& Zabludoff(2000)]{2000ApJ...545..129K} 
Keeton, C.~R., Christlein, D., \& Zabludoff, A.~I.\ 2000, ApJ, 545, 129

\bibitem[Keeton(2002)]{2002ApJ...575L...1K} Keeton, C.~R.\ 2002, ApJL, 575, L1

\bibitem[Kinney et al.(1996)]{1996ApJ...467...38K} Kinney, A.~L., Calzetti, 
D., Bohlin, R.~C., McQuade, K., Storchi-Bergmann, T., \& Schmitt, H.~R.\ 
1996, ApJ, 467, 38

\bibitem[Kneib et al.(1998)]{1998A&A...329..827K} Kneib, J.-P., Alloin, D., 
Mellier, Y., Guilloteau, S., Barvainis, R., \& Antonucci, R.\ 1998, A\&A, 
329, 827

\bibitem[Kneib, Cohen, \& Hjorth(2000)]{2000ApJ...544L..35K} Kneib, J., 
Cohen, J.~G., \& Hjorth, J.\ 2000, ApJ, 544, L35

\bibitem[Kobayashi, Terada, Goto, \& Tokunaga(2002)]{2002ApJ...569..676K} 
Kobayashi, N., Terada, H., Goto, M., \& Tokunaga, A.\ 2002, ApJ, 569, 676

\bibitem{K1992} Kochanek, C.S., 1992, ApJ, 384, 1

\bibitem{K1993} Kochanek, C.S., 1993, ApJ, 419, 12

\bibitem{K1994} Kochanek, C.S., 1994, ApJ, 436, 56

\bibitem{K1998} Kochanek, C.S, Falco, E.E., Impey, C.D., Leh\'{a}r, J., McLeod, B.A., Rix, H.-W., 1998, in After the Dark Ages: When Galaxies Where Young (AIP), ed. S. Holt \& E. Smith, p. 163

\bibitem[Kochanek et al.(2000)]{2000ApJ...543..131K} Kochanek, C.~S.~et 
al.\ 2000, ApJ, 543, 131

\bibitem[Koopmans et al.(1999)]{1999MNRAS.303..727K} Koopmans, L.~V.~E.~et 
al.\ 1999, MNRAS, 303, 727

\bibitem[Koopmans \& Treu(2002)]{2002ApJ...568L...5K} Koopmans, 
L.~;.~V.~E.~\& Treu, T.\ 2002, ApJL, 568, L5

\bibitem[Kormendy \& Djorgovski(1989)]{1989ARA&A..27..235K} Kormendy, J.~\& 
Djorgovski, S.\ 1989, ARA\&A, 27, 235

\bibitem[Lacy, Rawlings, \& Serjeant(1998)]{1998MNRAS.299.1220L} Lacy, M., 
Rawlings, S., \& Serjeant, S.\ 1998, MNRAS, 299, 1220

\bibitem[Lacy et al.(2002)]{2002AJ....123.2925L} Lacy, M., Gregg, M., 
Becker, R.~H., White, R.~L., Glikman, E., Helfand, D., \& Winn, J.~N.\ 
2002, AJ, 123, 2925

\bibitem[Langston et al.(1989)]{1989AJ.....97.1283L} Langston, G.~I.~et 
al.\ 1989, AJ, 97, 1283

\bibitem[Leh{\' a}r et al.(1993)]{1993AJ....105..847L} Leh{\' a}r, J., Langston, 
G.~I., Silber, A., Lawrence, C.~R., \& Burke, B.~F.\ 1993, AJ, 105, 847

\bibitem[Leh{\' a}r et al.(2000)]{2000ApJ...536..584L} Leh{\' a}r, J.~et 
al.\ 2000, ApJ, 536, 584

\bibitem[Leh{\' a}r et al.(2001)]{2001ApJ...547...60L} Leh{\' a}r, J., 
Buchalter, A., McMahon, R.~G., Kochanek, C.~S., \& Muxlow, T.~W.~B.\ 2001, 
ApJ, 547, 60

\bibitem[Lidman et al.(2000)]{2000A&A...364L..62L} Lidman, C., Courbin, F., 
Kneib, J.-P., Golse, G., Castander, F., \& Soucail, G.\ 2000, A\&A, 364, L62

\bibitem[Lin et al.(1999)]{1999ApJ...518..533L} Lin, H., Yee, H.~K.~C., 
Carlberg, R.~G., Morris, S.~L., Sawicki, M., Patton, D.~R., Wirth, G., \& 
Shepherd, C.~W.\ 1999, ApJ, 518, 533

\bibitem[Lubin et al.(2000)]{2000AJ....119..451L} Lubin, L.~M., Fassnacht, 
C.~D., Readhead, A.~C.~S., Blandford, R.~D., \& Kundi{\' c}, T.\ 2000, AJ, 
119, 451

\bibitem[Madgwick et al.(2002)]{2002MNRAS.333..133M} Madgwick, D.~S.~et 
al.\ 2002, MNRAS, 333, 133

\bibitem[Magain et al.(1988)]{1988Natur.334..325M} Magain, P., Surdej, J., 
Swings, J.-P., Borgeest, U., \& Kayser, R.\ 1988, Nature, 334, 325

\bibitem{MRT97} Malhotra, S., Rhoads, J.E., Turner, E.L. 1997,

\bibitem{MR93} Maoz, D., Rix, H.-W. 1993, ApJ, 416, 425

\bibitem[Marlow et al.(1999)]{1999AJ....118..654M} Marlow, D.~R.~et al.\ 
1999, AJ, 118, 654

\bibitem[Marlow et al.(2001)]{2001AJ....121..619M} Marlow, D.~R.~et al.\ 
2001, AJ, 121, 619

\bibitem[Martel, Premadi, \& Matzner(2002)]{2002ApJ...570...17M} Martel, 
H., Premadi, P., \& Matzner, R.\ 2002, ApJ, 570, 17

\bibitem{Morgan1999} Morgan, N. D.,  Dressler, A., Maza, J., Schechter, P. L., Winn, J. N., 1999, AJ, 118, 1444

\bibitem[Morgan et al.(2000)]{2000AJ....119.1083M} Morgan, N.~D.~et al.\ 
2000, AJ, 119, 1083

\bibitem[Morgan et al.(2001)]{2001AJ....121..611M} Morgan, N.~D., Becker, 
R.~H., Gregg, M.~D., Schechter, P.~L., \& White, R.~L.\ 2001, AJ, 121, 611

\bibitem[Mortlock \& Webster(2001)]{2001MNRAS.321..629M} Mortlock, D.~J.~\& 
Webster, R.~L.\ 2001, MNRAS, 321, 629

\bibitem{M1999} Munoz, J. A., Falco, E. E., Kochanek, C. S., Leh{\' a}r, J., McLeod, B. A., Impey, C. D., Rix, H.-W., Peng, C. Y., 1999, Ap\&SS, 263, 51

\bibitem[Mu{\~ n}oz et al.(2001)]{2001ApJ...546..769M} Mu{\~ n}oz, J.~A.~et 
al.\ 2001, ApJ, 546, 769

\bibitem[Myers et al.(1999)]{1999AJ....117.2565M} Myers, S.~T.~et al.\ 
1999, AJ, 117, 2565

\bibitem{NB1996} Narayan, R., \& Bartelmann, M., 1996, In: ``Lectures on Gravitational Lensing'', astro-ph/9606001

\bibitem[O'Dea et al.(1992)]{1992AJ....104.1320O} O'Dea, C.~P., Baum, 
S.~A., Stanghellini, C., Dey, A., van Breugel, W., Deustua, S., \& Smith, 
E.~P.\ 1992, AJ, 104, 1320

\bibitem{PR2002} Peebles, P.J.E., Ratra, B., 2002, astro-ph/0207347

\bibitem[Peng et al.(1999)]{1999ApJ...524..572P} Peng, C.~Y.~et al.\ 1999, ApJ, 524, 572

\bibitem[Perlmutter et al.(1999)]{1999ApJ...517..565P} Perlmutter, S.~et 
al.\ 1999, ApJ, 517, 565

\bibitem{P2000} Phillips, P.M., et al., 2000, 319, L7

\bibitem[Postman \& Geller(1984)]{1984ApJ...281...95P} Postman, M.~\& 
Geller, M.~J.\ 1984, ApJ, 281, 95

\bibitem[Premadi, Martel, Matzner, \& Futamase(2001)]{2001ApJS..135....7P} 
Premadi, P., Martel, H., Matzner, R., \& Futamase, T.\ 2001, ApJS, 135, 7

\bibitem[Press, Teukolsky, Vetterling, \& 
Flannery(1992)]{1992nrfa.book.....P} Press, W.~H., Teukolsky, S.~A., 
Vetterling, W.~T., \& Flannery, B.~P.\ 1992, Cambridge: University Press, 
|c1992, 2nd ed.,


\bibitem{R1999} Ratnatunga, K.U., Griffiths, R.E., Ostrander, E.J., 1999, AJ, 117, 2010

\bibitem[Reimers et al.(2002)]{2002A&A...382L..26R} Reimers, D., Hagen, 
H.-J., Baade, R., Lopez, S., \& Tytler, D.\ 2002, A\&A, 382, L26

\bibitem[Riess et al.(1998)]{1998AJ....116.1009R} Riess, A.~G.~et al.\ 
1998, AJ, 116, 1009

\bibitem[Rix, Maoz, Turner, \& Fukugita(1994)]{1994ApJ...435...49R} Rix, 
H., Maoz, D., Turner, E.~L., \& Fukugita, M.\ 1994, ApJ, 435, 49

\bibitem[Rix et al.(1997)]{1997ApJ...488..702R} Rix, H., de Zeeuw, P.~T., 
Cretton, N., van der Marel, R.~P., \& Carollo, C.~M.\ 1997, ApJ, 488, 702

\bibitem[Rusin \& Ma(2001)]{2001ApJ...549L..33R} Rusin, D.~\& Ma, C.\ 2001, ApJL, 549, L33

\bibitem[Rusin et al.(2001a)]{2001AJ....122..591R} Rusin, D.~et al.\ 2001a, 
AJ, 122, 591

\bibitem[Rusin et al.(2001b)]{2001ApJ...557..594R} Rusin, D.~et al.\ 2001b, 
ApJ, 557, 594

\bibitem[Rusin et al.(2002a)]{2002MNRAS.330..205R} Rusin, D., Norbury, M., 
Biggs, A.~D., Marlow, D.~R., Jackson, N.~J., Browne, I.~W.~A., Wilkinson, 
P.~N., \& Myers, S.~T.\ 2002a, MNRAS, 330, 205

\bibitem{Rusin2002} Rusin, D, et al. 2002b, accepted to ApJ, astro-ph/0211229

\bibitem[Schade et al.(1999)]{1999ApJ...525...31S} Schade, D.~et al.\ 1999, ApJ, 525, 31

\bibitem[Schechter(1976)]{1976ApJ...203..297S} Schechter, P.\ 1976, ApJ, 
203, 297

\bibitem[Schechter et al.(1998)]{1998AJ....115.1371S} Schechter, P.~L., 
Gregg, M.~D., Becker, R.~H., Helfand, D.~J., \& White, R.~L.\ 1998, AJ, 
115, 1371

\bibitem[Schlegel, Finkbeiner, \& Davis(1998)]{1998ApJ...500..525S} 
Schlegel, D.~J., Finkbeiner, D.~P., \& Davis, M.\ 1998, ApJ, 500, 525


\bibitem{Saho1995} Shao, J., Tu., P, 1995, in: The Jackknife and Bootstrap, Springer series in statistics, Springer Verlag

\bibitem[Siemiginowska et al.(1998)]{1998ApJ...503..118S} Siemiginowska, 
A., Bechtold, J., Aldcroft, T.~L., McLeod, K.~K., \& Keeton, C.~R.\ 1998, 
ApJ, 503, 118

\bibitem[Small, Sargent, \& Steidel(1997)]{1997AJ....114.2254S} Small, 
T.~A., Sargent, W.~L.~W., \& Steidel, C.~C.\ 1997, AJ, 114, 2254

\bibitem[Surdej et al.(1997)]{1997A&A...327L...1S} Surdej, J., Claeskens, 
J.-F., Remy, M., Refsdal, S., Pirenne, B., Prieto, A., \& Vanderriest, C.\ 
1997, A\&A, 327, L1

\bibitem[Sykes et al.(1998)]{1998MNRAS.301..310S} Sykes, C.~M.~et al.\ 
1998, MNRAS, 301, 310

\bibitem[Tonry(1998)]{1998AJ....115....1T} Tonry, J.~L.\ 1998, AJ, 115, 1

\bibitem[Tonry \& Kochanek(1999)]{1999AJ....117.2034T} Tonry, J.~L.~\& 
Kochanek, C.~S.\ 1999, AJ, 117, 2034

\bibitem[Tonry \& Kochanek(2000)]{2000AJ....119.1078T} Tonry, J.~L.~\& 
Kochanek, C.~S.\ 2000, AJ, 119, 1078

\bibitem{TK2002} Treu, T., Koopmans, L.~V.~E., 2002, accepted to ApJ, astro-ph/0202342

\bibitem[Turner(2002)]{2002ApJ...576L.101T} Turner, M.~S.\ 2002, ApJL, 576, L101

\bibitem{TOG84} Turner, E.L., Ostriker, J.P., Gott III, R. 1984, ApJ, 284, 1

\bibitem[van Dokkum, Franx, Kelson, \& 
Illingworth(1998)]{1998ApJ...504L..17V} van Dokkum, P.~G., Franx, M., 
Kelson, D.~D., \& Illingworth, G.~D.\ 1998, ApJL, 504, L17 

\bibitem[van Dokkum, Franx, Kelson, \& 
Illingworth(2001)]{2001ApJ...553L..39V} van Dokkum, P.~G., Franx, M., 
Kelson, D.~D., \& Illingworth, G.~D.\ 2001, ApJL, 553, L39

\bibitem[de Vaucouleurs \& Olson(1982)]{1982ApJ...256..346D} de 
Vaucouleurs, G.~\& Olson, D.~W.\ 1982, ApJ, 256, 346

\bibitem[Walsh, Carswell, \& Weymann(1979)]{1979Natur.279..381W} Walsh, D., 
Carswell, R.~F., \& Weymann, R.~J.\ 1979, Nature, 279, 381

\bibitem{W1996} Warren, S.J., Hewett, P.C., Lewis, G.F., Moller, P., Iovino, A., Shaver, P.A., 1996, MNRAS, 278, 139

\bibitem[Wiklind \& Combes(1995)]{1995A&A...299..382W} Wiklind, T.~\& 
Combes, F.\ 1995, A\&A, 299, 382

\bibitem[Wiklind \& Combes(1996)]{1996Natur.379..139W} Wiklind, T.~\& 
Combes, F.\ 1996, Nature, 379, 139

\bibitem[Winn et al.(2000)]{2000AJ....120.2868W} Winn, J.~N.~et al.\ 2000, 
AJ, 120, 2868

\bibitem[Winn et al.(2001)]{2001AJ....121.1223W} Winn, J.~N., Hewitt, 
J.~N., Patnaik, A.~R., Schechter, P.~L., Schommer, R.~A., L{\' o}pez, S., 
Maza, J.~;, \& Wachter, S.\ 2001, AJ, 121, 1223

\bibitem[Winn et al.(2002a)]{2002AJ....123...10W} Winn, J.~N.~et al.\ 2002a, AJ, 123, 10

\bibitem[Winn et al.(2002b)]{2002ApJ...564..143W} Winn, J.~N., Lovell, 
J.~E.~J., Chen, H., Fletcher, A.~;., Hewitt, J.~N., Patnaik, A.~R., \& 
Schechter, P.~L.\ 2002b, ApJ, 564, 143

\bibitem{Wis1999} Wisotzki, L., Christlieb, N., Liu, M.~C., Maza, J., Morgan, N.~D., Schechter, P.~L., 1999, A\&A, 348, L41

\bibitem[Wisotzki et al.(2002)]{2002A&A...395...17W} Wisotzki, L., 
Schechter, P.~L., Bradt, H.~V., Heinm{\" u}ller, J., \& Reimers, D.\ 2002, A\&A, 395, 17

\bibitem[White \& Davis(1996)]{1996AAS...189.4104W} White, R.~E.~\& Davis, 
D.~S.\ 1996, American Astronomical Society Meeting, 28, 1323

\bibitem[Zabludoff \& Mulchaey(1998)]{1998ApJ...496...39Z} Zabludoff, 
A.~I.~\& Mulchaey, J.~S.\ 1998, ApJ, 496, 39

\end{thebibliography}
\end{document}